\documentclass[acmsmall]{acmart}
\usepackage{csquotes}
\usepackage{multirow}

\AtBeginDocument{%
 }

\setcopyright{rightsretained}
\acmJournal{PACMHCI}
\acmYear{2024} \acmVolume{8} \acmNumber{CSCW1} \acmArticle{171} \acmMonth{4} \acmPrice{}\acmDOI{10.1145/3641010}

\makeatletter
\gdef\@copyrightpermission{
  \begin{minipage}{0.2\columnwidth}
   \href{https://creativecommons.org/licenses/by/4.0/}{\includegraphics[width=0.90\textwidth]{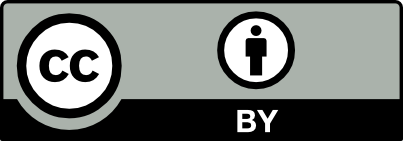}}
  \end{minipage}\hfill
  \begin{minipage}{0.8\columnwidth}
   \href{https://creativecommons.org/licenses/by/4.0/}{This work is licensed under a Creative Commons Attribution International 4.0 License.}
  \end{minipage}
  \vspace{5pt}
}
\makeatother

\usepackage{graphics}
\usepackage{tikz}
\usepackage{subcaption}
\usepackage[normalem]{ulem}
\usepackage{tcolorbox}
\usetikzlibrary{shapes.geometric}
\usepackage{multirow}
\usepackage{longtable}

\usepackage{tabularray}
\usepackage{rotating}
\usepackage{pdflscape}
\usepackage{array}

\newcommand{\add}[1]{\textcolor{black}{\noindent{#1}}}

\begin{document}

\title{Misinformation as a Harm: Structured Approaches for Fact-Checking Prioritization}

\author{Connie Moon Sehat}
\affiliation{%
  \institution{Hacks/Hackers}
  \country{USA}
}
\email{connie@hackshackers.com}

\author{Ryan Li}
\affiliation{%
  \institution{Stanford University}
  \country{USA}}
\email{lansong@stanford.edu}

\author{Peipei Nie}
\affiliation{%
 \institution{University of Washington}
 \country{USA}}
\email{niep@cs.washington.edu}

\author{Tarunima Prabhakar}
\affiliation{%
  \institution{Tattle Civic Technologies}
  \country{India}}
\email{tarunima@tattle.co.in}

\author{Amy X. Zhang}
%\authornote{Corresponding author}
\affiliation{%
  \institution{University of Washington}
  \country{USA}}
\email{axz@cs.uw.edu}

%\renewcommand{\shortauthors}{Sehat et al.}
\renewcommand{\shortauthors}{Connie Moon Sehat et al.}

\begin{abstract}
In this work, we examine how fact-checkers prioritize which claims to fact-check and what tools may assist them in their efforts. Through a series of interviews with 23 professional fact-checkers from around the world, we validate that harm assessment is a central component of how fact-checkers triage their work. We also clarify the processes behind fact-checking prioritization, finding that they are typically ad hoc, and gather suggestions for tools that could help with these processes. 

To address the needs articulated by fact-checkers, we present a structured framework of questions to help fact-checkers negotiate the priority of claims through assessing potential harms. Our FABLE Framework of Misinformation Harms incorporates five dimensions of magnitude---\textit{(social) Fragmentation}, \textit{Actionability}, \textit{Believability}, \textit{Likelihood of spread}, and \textit{Exploitativeness}---that can help determine the potential urgency of a specific message or claim when considering misinformation as harm. The result is a practical and conceptual tool to support fact-checkers and others as they make strategic decisions to prioritize their efforts. We conclude with a discussion of computational approaches to support structured prioritization, as well as applications beyond fact-checking to content moderation and curation.

\end{abstract}




% For Articles V8cscw010-V8cscw158 (updated on 2/1/24 due to editors' miscommunication), use:
%\received{July 2023}
%\received[revised]{October 2023}
%\received[accepted]{December 2023} 

% For Articles V8cscw159-V8cscw192, use:
\received{July 2023}
\received[revised]{October 2023}
\received[accepted]{November 2023}

\maketitle

Online misinformation is a major challenge for societies today. Beliefs in false claims about science, such as vaccine misinformation, can lead people to engage in harmful behavior that risks their own health. Such misinformed beliefs can also defeat public health measures that rely on collective compliance to protect society’s most vulnerable \cite{Schuster_Eskola_Duclos_2015, Loomba_deFigueiredo_Piatek_deGraaf_Larson_2021, Bozzola_Spina_Russo_Bozzola_Corsello_Villani_2018, Pandey_Galvani_2023}. Similarly, a belief in inaccurate or misleading narratives about vote-rigging or other supposed election interference can lower the public's trust in democratic institutions, and in turn affect the level of participation in political activities such as voting, interfere with the peaceful transition of power, and even stoke violent rhetoric and action \cite{Nisbet_Mortenson_Li_2021, Riley_2022, Bennett_Livingston_2018, Jan6Report_2022}. 

Fact-checking is a critical strategy when addressing misinformation~\cite{walter2020fact,Stencel_Ryan_Luther_2022}. Fact-checking supports individual readers who seek good information, and also supports content moderation and labeling initiatives on large scale platforms~\cite{zhang2021effects,clayton2020real}. However, fact-checking is laborious. The fact-checking process includes investigating claims, collecting convincing evidence that such claims are false or misleading, and then sharing that evidence out. With torrential volumes of user-generated content created daily, it is impossible to fact-check every new article, post, message, or claim.
As a result, fact-checkers tasked with addressing online misinformation must prioritize what they choose to tackle. Given that prioritization is unavoidable, how should fact-checking efforts to combat misinformation prioritize what content to tackle? Can the prioritization be systematized? Can a systematic process also reflect the priorities and desires of fact-checkers? 

One way forward is through \textit{harm assessment}. Taking the approach that misinformation could be treated as a harm opens up a fruitful line of inquiry, as the perspective of misinformation as a harm aligns with the motivations of fact-checkers. Like the journalism field out of which it was born, fact-checking has at its heart altruistic ideals such as holding power accountable and helping the public to achieve informed decision-making \cite{Graves_Nyhan_Reifler_2016}. In addition, while all misinformation is harmful to some degree, not all misinformation is equally harmful, making harm assessment a potentially useful component of prioritization. 

Through a series of interviews with 23 professional fact-checkers from around the world, we validated that harm assessment is a central component of how fact-checkers triage their work. We gained an understanding of how fact-checkers determine harm, including what they look for, how and when they incorporate harm assessment into their process, and the other factors considered, when prioritizing what to fact-check. We also learned about the role of tools, existing and proposed, that could support this process. In summary, we sought the answers to the following research questions: 
\begin{itemize}
\item RQ1: According to fact-checkers, what aspects of misinformation create urgency or importance?
\item RQ2: How do fact-checkers decide what to fact-check and what tools could improve their processes of prioritization?
%\item RQ3: What tools would be helpful to fact-checkers with regard to prioritization?
\end{itemize}

From our interviews, we discover that fact-checkers take many considerations into account when prioritizing. Key among their concerns is the potential harmfulness of a claim (particularly when it is physical), the claim's likelihood of spread or virality, and the potential impact of a fact-check. We also find that fact-checking processes, overall, are still young and not standardized. Fact-checkers typically take a relatively ad hoc approach to prioritization, using individual judgment and case-by-base discussion with others.  Regarding tools, fact-checkers desire features that can help ease their work or speed up their processes, as well as tools that help them assess the potential harmful impact of misinformation in ways that are sensitive to local context.

Drawing on these findings, we present a novel \textit{misinformation harms framework} to enable fact-checkers with prioritization in a more structured fashion. 
% We begin by making the observation that while all misinformation is harmful to some degree, not all misinformation is equally harmful. 
Following a literature review, workshops with fact-checkers and other misinformation experts, and an incorporation of the interview findings, we developed dimensions of analysis to help prioritize fact-checking efforts in the format of a questionnaire. 
Using a draft of the taxonomy and accompanying questionnaire, we received feedback from 4 additional professional fact-checkers, and iterated again on the dimensions and questions.

Our \add{\textbf{FABLE Framework of Misinformation Harms} incorporates five dimensions of magnitude---\textit{(social) Fragmentation}, \textit{Actionability}, \textit{Believability},  \textit{Likelihood of spread}, and \textit{Exploitativeness}}---that can help determine the potential urgency of a specific message or post when considering misinformation as harm. 
The framework, and its questions, are intended as both conceptual and practical tools. Based on the desires and perspectives of fact-checkers, the framework may support them (as well as content moderators, peer correction efforts, and other initiatives) as they make strategic decisions when prioritizing their efforts to respond to misinformation that is spreading.  We discuss ways our framework and questionnaire could be used within misinformation response practice and also discuss design implications for tools to support misinformation response.
\begin{figure}
    \centering
    \includegraphics[width=0.90\textwidth]{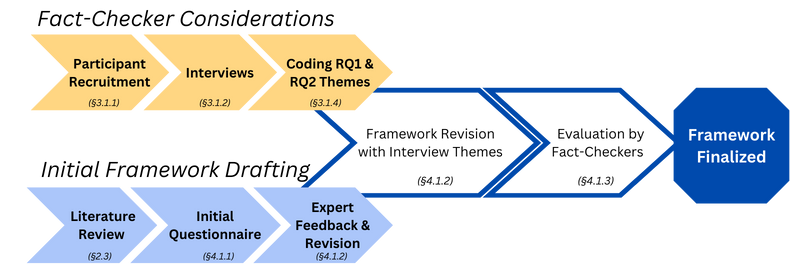}
    \caption{\add{Methodological Process for the development of the FABLE Framework of Misinformation Harms.}}
    \label{fig:process}
\end{figure}

\section{Related Work}

\subsection{Harms of Misinformation}
% - Misinformation and disinformation around the COVID-19 pandemic or elections around the world can have impacts ranging from peoples’ health to the stability of democratic institutions.
 
%- impact of misinformation

%- more misinformation/greater spread = bad for society

The question of whether false information is harmful itself is perhaps as old as human society. Do all untruths damage others? What if they are intended to prevent harm, such as white lies? How about misleading statements or omissions of fact? Philosophers and theologians have engaged in questions around truth and falsehood for millennia. There are moral dilemmas behind lying at an individual and social level, even for just the ``harmless white lie'' \cite{bok_lying_2011}. For the last two decades, concerns from journalists, political scientists, cognitive psychologists, and other research communities about whether we are in a ``post-truth'' society have added to this conversation. Even as finer points are debated, scholars acknowledge the deleterious effects of lying upon interpersonal trust, overall sociability, and even the ability to hope \cite{Keyes_2004, Lewandowsky_Ecker_Cook_2017, Franklin_McNair_2017, Iyengar_Massey_2019, Farkas_Schou_2019, Snyder_2002}. 

Moreover, harm itself is a complex social and legal concept that involves a process of clarifying, or classifying, its relative degrees of effect and corresponding proscriptions or punishments. In online realms a wide range of socially undesirable content, such as harassment, child exploitation and narratives leading to self-harm, can be characterized as harmful \cite{Sehat_Lalani_2021}. Definitions of harm, like those related to truth, can be socially dependent and may involve the evaluation of multiple incidents over time, making context and nuance critical. And, at least within democracies, attempts to assess relative degrees of harm must maintain the fine balance against diminishing other human rights regarding the freedom of speech and conscience and rights to free assembly \cite{Feinberg_1987, Brettschneider_2012}. 

Practically speaking, existing practices for addressing potentially harmful content face the challenge of triage. In particular, content moderators and fact-checkers must contend with a tidal wave of content shared via the internet. Even with automated AI tools to help remove spam, platforms such as Facebook still had over 3 million pieces of flagged content every day in 2020 \cite{Barrett_2020}. Reports include people working to review between 25 to 100 pieces of content every hour \cite{Barrett_2020, shead_tiktok_2020}.  More examples of this scale and challenge can be seen in reports from other platforms such as YouTube or Reddit \cite{YouTube_Community_Transparency_Report, Reddit_Transparency_2021}. 

Bringing a structured harm assessment to misinformation, then, means to prioritize according to its potential harmful effect---while all misinformation is harmful to some degree, not all misinformation is equally harmful. As an example, compare a hoax about a celebrity death versus the false claim about toxic seeds that supposedly provide COVID-19 immunity.\footnote{See for example \url{https://en.wikipedia.org/wiki/Death_hoax} in contrast to \cite{Hindu_Datura_2020}} For a variety of reasons, it may be more urgent to try to combat the latter example of misinformation.

\subsection{Fact-checking Practices}
%- Faced with this challenge, fact-checking has entered a new era.
 
Online platforms that focus on user-generated content have been increasingly exploring ways to scale content review to support safe and accountable exchanges of information, in order to match the pace of online distribution. This is one reason for an explosion of growth in the field of fact-checking, which originally grew out of magazine journalism in the earlier half of the twentieth century \cite{Graves_2016}. Fact-checking has emerged as a distinct profession, independent of journalistic training or traditional journalism channels such as newspapers---for example, a 2022 census counted 391 fact-checking organizations in contrast to 186 in 2016 \cite{Stencel_Ryan_Luther_2022}. The demand for accurate information in different contexts, such as elections and health, continues to change the nature of fact-checking work \cite{Fischer_2020, Siwakoti_Yadav_Bariletto_Zanotti_Erdogdu_Shapiro_2021}.

To understand how well fact-checking organizations have adapted to these demands, researchers have unpacked the fact-checking process by revealing the human and technological infrastructures that support and shape fact-checking work \cite{juneja2022human}, and surfaced a pipeline of practices fragmented across disparate tools that lack integration \cite{micallef2022true}. Juneja and Mitra find that fact-checking is more than one-off debunking of misleading claims; it also involves long-term advocacy work to improve the information ecosystem \cite{juneja2022human}. This finding resonates with Micallef et al.'s work, where participants mentioned that they contribute to the information ecosystem to facilitate the creation of a balanced public sphere for discussing issues \cite{micallef2022true}. \add{Also worthy of consideration are the developments in less professionalized contexts. Work in collaborative and crowdsourced approaches to fact-checking reveal similar concerns and discussions, where the processes of this voluntary work again go well beyond simple debunking \cite{hehe2022, yasseri_menczer2023}}.

With regard to increasing demands for scaling fact-checking practices through automated tools, researchers find that the largely manual and labor intensive nature of current fact-checking practices is a barrier to scale \cite{micallef2022true}. This resonates with other work, which argues that quality data is essential not only for developing AI-based automated tools but also for investigating claims \cite{juneja2022human}.
Given the difficulty of scaling up fact-checking, our work considers the question of triage, in which time-consuming human efforts are assigned to areas that may have the greatest impact.
We also shed additional light into the practices, struggles, and needs of fact-checkers by interviewing a diverse set of fact-checkers about how they approach prioritization in their work.

% - how fact-checkers currently do fact-checking
% - practices/priorities
% - what tools they use
% - current problems with fact-checking, limitations?

\subsection{Structured Harm and Misinformation Assessments}
There are a number of works that have attempted to assess harm or misinformation in structured ways. Examples can be found in cybersecurity literature thinking about harms, such as economic, social, and even reputational harms. Other frameworks coming from a content moderation perspective focus on either misinformation or harm, with the weight of harm- versus misinformation-related definitions depending on the purpose of the classification. 
Here we present an overview of many taxonomies in this space, while in Section~\ref{sec:framework-dev}, we dive into a few relevant frameworks in more detail to discuss the dimensions we developed in our FABLE misinformation harms framework. 

 \subsubsection{Misinformation-oriented taxonomies}  
 We first describe taxonomies of misinformation, most of which focus on categorizing and characterizing different forms of misinformation according to their content or goal.
 Fitzgerald et al. in 1997 was probably one of the earliest approaches to evaluate online misinformation, identifying 10 types of online misinformation \cite{fitzgerald1997misinformation}.
 A more recent taxonomy identified by the First Draft organization, with authors Wardle and Derakhshan, has become widely adopted \cite{ wardle_fake_2017, Wardle_Derakhshan_2017}. The main focus of their conceptual definitions was to characterize false and misleading information, though their additional category of `mal-information'--- factual information used to inflict harm--- does intersect with harms-related considerations. 
  Nakamura et al. followed this work to present a dataset of Reddit posts classified under Wardle's seven types of fake news \cite{nakamura-etal-2020-fakeddit}.
 Other attempts include McCright and Dunlap's, which proposes four high level types---truthiness, bullshit, systemic lies, and shock-and-chaos \cite{mccright2017combatting}. 

 Researchers have characterized  misinformation using a range of lenses and methods to develop alternative taxonomies.
 Jiang et al. focused on misinformation stories and came up with 10 categories by studying the archived fact-checks from Snopes.com \cite{jiang2021structurizing}. 
 Charquero-Ballester et al. categorized COVID-19 misinformation into six types, through a focus on their claims or narratives, and compared the emotional valence among the different types \cite{charquero2021different}.
 % Ruokolainen et al. established a 4-dimension taxonomy of misinformation encountered by asylum seekers \cite{ruokolainen2020conceptualising}.
 In contrast, Brennen et al. focused on identifying the common formats, sources, and claims of COVID-19 misinformation \cite{brennen2020types}. 
   Psychological studies of misinformation have also differentiated between neutral versus non-neutral misinformation \cite{morgan2013misinformation}, and contradictory versus additive misinformation \cite{moore2016use}.

 \subsubsection{Harm-oriented taxonomies} 
 Separately, there are  taxonomies focused on online harms, broadly construed.
 Agrafiotis et al. have defined a taxonomy of harms from a cybersecurity perspective, where dimensions of harm defined include economic, social, and reputational harms \cite{agrafiotis_cyber_2016, agrafiotis_taxonomy_2018}. 
 Recent works have also studied the types and targets of offensive online content or online hate for purposes including content moderation. Zampieri et al. presented a three-layer annotation scheme for detecting offensive online content \cite{zampieri2019predicting}. The paper labels offensive content by whether it is a targeted insult or untargeted profanity, and whether it is targeting an individual, a group, or something else (e.g., an organization, a situation, an event, or an issue). Thomas et al.  presented a taxonomy of seven categories of online threats and attacks (toxic content, content leakage, overloading, false reporting, impersonation, surveillance, and lockout and control) \cite{thomas2021sok}. And Salminen et al. created a granular taxonomy for hateful online comments, describing 4 types of offensive language, 9 types of targets, and 16 types of sub-targets \cite{salminen2018anatomy}.  Furthermore, Tran et al.  specified 15 different types of harms related to misinformation, which were distilled from papers, and then focused on their presence within two different humanitarian crisis scenarios \cite{tran_investigation_2020}. Finally, most major platforms publish community guidelines that outline specific categories of objectionable content, and some have released detailed annotation guidelines that they provide to their paid content moderation staff.
 
 \subsubsection{Urgency- or severity-related taxonomies of misinformation harms}
 Most relevant to our work are taxonomies that bring together an evaluation of harms with the context of misinformation, and that have dimensions that can speak to a degree of urgency or severity.
 First, Scheuerman et al. established a framework of severity for harmful online content by considering approaches of severity from legal, law enforcement, and health professional perspectives. Taking 66 types of violations across 11 different social media platforms, the authors further refined the classifications to 20 categories; however the consideration of misinformation only appears within the single category of Coordinating Scams and Political Attacks \cite{scheuerman_framework_2021}. More closely related are non-profit organization FullFact's white papers, which have attempted to establish a framework of severity around `information incidents,' or large-scale public incidents where the coordination could better occur across organizations and institutions. Their approach is less about incidents regarding individuals (unless public figures), and more concerned around issues such as public health and elections; misinformation is clearly of concern in these issues \cite{fullfact_towards_2020}. Finally, though not explicitly incorporating severity, Mirza et al. has recently proposed a cybersecurity-inspired framework that characterizes disinformation threats by four dimensions: threat actors, attack patterns, attack channels, and target audience \cite{EURECOM+7114}.

As can be seen, while there are many frameworks relating to either misinformation or harms, there are few that focus specifically upon the harms resulting from misinformation and the potential levels of severity or urgency that may arise. In addition \add{to addressing this gap}, our work goes a step further to provide practically applicable worksheets that can be used by fact-checkers to systematize their own processes around prioritization.

\section{Interview Study}

When embarking on research around misinformation as a harm, we found few public resources or research discussing how fact-checkers prioritize their work. In order to validate our own perception of harm as a main factor in assessing the urgency of addressing misinformation, we decided to gather the direct experiences and perceptions of fact-checkers. We conducted semi-structured interviews with professional fact-checkers from certified fact-checking organizations that explored various aspects of their work---not only how much notions of potential harm motivated their own processes, but also their processes of triage. \add{We focused on professional fact-checkers compared to volunteers since the contractual nature of their work makes the question of triage more pressing.} In addition, we wanted to know more about how their work might be made easier with new or better tools.

\subsection{Method}
This section describes participant recruitment, our study protocol, characteristics of our participant sample, and how we conducted analysis \add{(\autoref{fig:process})}. All study procedures were reviewed and deemed exempt by University of Washington's Institutional Review Board (IRB).

\subsubsection{Participant Recruitment}
We recruited a total of 23 participants with fact-checking experience who were actively working within a fact-checking organization or team. 12 participants were recruited from personal connections of our collaborators. 11 participants were recruited through cold contact---in September 2021, we sent 53 emails to fact-checking organizations certified by the International Fact-Checking Network (IFCN). We manually collected these email addresses from the IFCN website. In our email, we provided a web page describing the study, with the link at the end for people to sign up.
We selected participants based on details provided such as their organization and location, in order to have a diverse representation of fact-checking organizations.

% We then emailed participants and scheduled a time to meet with them. 

\subsubsection{Study Protocol}
We interviewed participants individually using a semi-structured protocol that covered the following themes: background information about the fact-checkers and their organizations, typical fact-checking processes, how they make decisions about prioritization, both as individuals and as part of an organization. We also asked how they considered harms when prioritizing, and what aspects of misinformation they look for when assessing harms. Finally, we asked about their use of or interest in computational or other tools to support prioritization. 
% Within the above themes, we covered characteristics of the harms of misinformation such as: selecting claims for fact-checking; collaborating with domain experts, journalists, and local citizens; using computation tools, etc. 
All interviews took place over a video conferencing software. Interviews lasted around 60 minutes each and were conducted in English. All participants were compensated with a \$35 gift card; however, four participants declined the compensation.

\begin{table}[]
\small
\begin{tabular}{|l|lll|}
 \hline
\textbf{\#} & \textbf{Gender} & \textbf{Country} & \textbf{Role} \\ \hline 
P1          & Male           & South Africa         & Research role          \\ \hline
P2          & Female         & Ukraine              & Editing role  \\ \hline
P3          & Male           & India                &  Organizational role               \\ \hline
P4          & Male           & India                & Organizational role              \\ \hline
P5          & Female         & South Africa         & Fact-checker          \\ \hline
P6          & Male           & Greece               & Editing role      \\ \hline
P7          & Male           & France               & Fact-checker         \\ \hline
P8          & Male           & India                & Fact-checking role      \\ \hline
P9          & Female         & Australia            & Editor           \\ \hline
P10         & Female         & [South American country]            & Organizational role        \\ \hline
P11         & Male           & United States        & Fact-checker        \\ \hline
P12         & Male           & [Asian country]               & Organizational role                   \\ \hline
P13         & Male           & Italy                & Organizational role                 \\ \hline
P14         & Female         & Sweden               & Organizational role     \\ \hline
P15         & Female         & France               &  Trainer    \\ \hline
P16         & Female         & Ukraine              & Editing role        \\ \hline
P17         & Male           & Poland               & Organizational role              \\ \hline
P18         & Male           & Brazil               & Organizational role       \\ \hline
P19         & Male           & Nigeria              & Editing role           \\ \hline
P20         & Male           & France               & Editing role     \\ \hline
P21         & Male           & France               & Organizational role             \\ \hline
P22         & Male           & Spain                & Coordinator             \\ \hline
P23         & Female         & India                & Organizational role      \\ \hline
 
\end{tabular}
\caption{Demographic details of interviewees for this study; countries redacted for de-identification.}
\label{table:participants}
\end{table}

%\begin{table}[]
%\small
%\begin{tabular}{|l|l|}
% \hline
%\textbf{Count} & \textbf{Organization Size} \\ \hline 
%1          & Small \\ \hline
%4          & Medium     \\ \hline
%18          & Large             \\ \hline
 
%\end{tabular}
%\caption{Organizational sizes of interviewees for this study.}
%\label{table:participants_size}
%\end{table}

\subsubsection{Participant Demographics}
Table 1 provides an overview of our participant sample. The 23 participants (8 women and 15 men) were from 15 countries covering Africa, Asia, Europe, Latin America, North America, and Oceania. The sample included fact-checkers from a mix of organizations, with 1 from small fact-checking organizations (3--6 employees), 4 from medium fact-checking organizations (7--12 employees), and 18 large fact-checking organizations (more than 12 employees). We omit the names of the fact-checking organizations in order to preserve the anonymity of our participants.

\subsubsection{Data collection and analysis}
We recorded all interviews with participant permission and transcribed them for analysis. We analyzed the transcripts using a Grounded Theory approach~\cite{strauss1997grounded}. This approach allowed common themes to emerge from the data in an inductive and interpretative manner. Specifically, we randomly selected three transcripts, and two of our authors then open coded them independently. During the open coding phase, the two authors coded the data on a sentence-by-sentence basis and created codes without initial hypotheses. They regularly came together to discuss and resolve disagreements on the codes. 
% Then they labeled each sentence with an underlying concept.
Subsequently, they examined the codes for similarities, removed the redundant codes, and created a codebook with definitions for each code. All the authors also reviewed the codebook and discussed the definitions and possible overlapping codes. Then, the two authors went on to split up the remaining transcripts and independently coded them, while continuing to discuss and iterate on the shared codebook with each other and the full team as new codes arose.
We reached theoretical saturation after analyzing 18 out of the 23 interviews as no new codes emerged after that point. \add{See Supplementary Materials for the full set of codes. These thematic codes resulted in the answers to our RQs, discussed below.}

% \section{Findings}

% [intro findings more -- All the findings below only represent fact-checkers and editors’ perspectives]

\subsection{RQ1: According to fact-checkers, what aspects of misinformation create urgency or importance?}
\label{sec:urgency}

We begin with a focus on how fact-checkers prioritize their review of inaccurate and contested claims.
Specifically, we dive into the aspects that many of our interviewees mentioned when they gauge the importance and urgency around misinformation.
We also describe how fact-checkers talked about how importance relates to the way misinformation plays out differently in different regions of the world.
%We identified two primary factors our fact-checkers mentioned making misinformation harmful: misinformation stirring physical and societal harm, and accumulative effects of misinformation.

\subsubsection{Whether the misinformation may lead to different types of harmful impact}

Interviewees often reflected upon or differentiated between two following types of harmful impact when considering urgency of misinformation. If a piece of misinformation can be argued to possibly cause negative physical or societal effects, which can at times be intertwined, then interviewees considered it important to address.

\begin{itemize}
    \item \textbf{Physical harm.} All participants reported that they would prioritize fact-checking misinformation that has potential for physical harm and considered it the biggest threat of misinformation. One form of physical harm that was repeatedly mentioned was the use of misinformation to provoke violent attacks by stoking outrage or calling for retribution against a group.
% Physical harm reported by our participants that can be directly caused by misinformation include violence and poor decision-making about personal health. 
% \begin{displayquote}
% ``\textit{I think the biggest threat is people making violent attacks on others, people making bad decisions about their health.}'' (P15)
% \end{displayquote}
% \begin{displayquote}
% “\textit{Often it'd be claim about some horrible thing that the other group would have done - some false claim. It can often be something about how they harm women, how they harm children - things that are very emotional. I felt often involving women or children, accusations of rape - things that are really meant to stoke really strong emotions, and that we see pretty often.” (P15)
% \end{displayquote}
The other kind of physical harm mentioned was misinformation that may lead people to make poor health choices.
Participants highlighted vaccine misinformation in particular as urgent because of its link to physical harm, as misleading information may convince people against getting a possibly life-saving vaccine: 

\begin{displayquote}
“\textit{Oftentimes when things are circulating a lot online, it spills into real life. The thing about a vaccine is also urgent for the same reason, because the more you read about the problem with the vaccine online, then you just don't want to get it.}”  (P9) 
\end{displayquote}

 % And then what's going to happen for society as a whole is that you are not going to have a high vaccination coverage, high enough to open up.

Some participants also justified the urgency of addressing certain misinformation using the reasoning of physical harm even if the effect was not necessarily direct or immediate. For instance, misinformation that casts doubt on climate change could cause significant physical harm as a second-order effect, as disbelief about climate change could lead to inaction on policy, which then leads to more climate-related deaths. \vspace{2mm}

\item \textbf{Societal harm.} 
Another form of harm that was highly referenced was societal harm, or harm that adversely impacts the cohesion and functioning of a society. For instance, a particularly prominent kind of misinformation that our participants considered urgent was election misinformation, which does not necessarily directly cause physical harm. 
However, similar to the case of climate misinformation, some participants still linked societal harm to physical harm as a second-order effect:

\begin{displayquote}
“\textit{It doesn't really kill you even if you believe Donald Trump has won, but it threatens democracy, which is the pillar of our society. Then you see people were organizing to go to the Capitol before January 6, right? They were talking about that on...all those forums. 
And then, there were people who died in the thing.}” (P9) 
\end{displayquote}

\end{itemize}

As can be seen, physical and societal harm can be deeply intertwined, and in many cases, interviewees felt that one will also imply the other. However, of the two, interviewees signaled the greater importance of physical harm. Not only did all our participants mention potential physical harm of misinformation as a priority, but individual interviewees highlighted its direct and immediate impacts as the underlying rationale for urgent address.

% One participant was worried that election misinformation would undermine our democracy, and the lack of trust in institutions can then lead to violence, as another example of physical harm being a second-order consequence of misinformation. 

%Another reason for participants to prioritize fact-checking misinformation that has potential for physical harm is that it can cause societal harm. A prominent case is that if rampant vaccine misinformation is left untreated, it can lead to vaccine hesitancy and lower rates of vaccination, which in turn can cause slower opening of the society. 

% Elaborate on what they meant by societal harm. Bring more clarity about how they think about societal harm. Link emotion/sensation into harm/action (actionalibility, what leads to action, what kind of actions are harm). Motivates development of our framework with org-wide framework. Included how many times. 12 out of 28.

\subsubsection{Characteristics of the misinformation content}

In addition to discussing the possible impacts of a piece of misinformation, our interviewees talked about specific ways that the misinformation itself was presented or framed that might make it more or less important to address. Most of the time, they described characteristics of the content that would make it more likely for the audience to believe and then act on the content, thus potentially leading to physical or societal harm, or share the content further, thus exposing more people to the misinformation.

\begin{itemize}
    \item \textbf{Emotional and sensational language.} Almost all participants mentioned the role that emotional language and sensationalism play in the virality of misinformation. People writing misinformation content will often use tactics to stoke emotions such as outrage with the goal of getting it shared further:

\begin{displayquote}
“\textit{I guess if it appeals to the emotions in a certain way, this is something that I see frequently, where the poster will say whatever claim, and ‘This should make you angry,’ and ‘I can't believe this is happening,’ things along those lines. Just framing misinformation in such a way that appeals to emotions I think makes it much more likely that this piece of misinformation will then get reshared.}” (P11)
\end{displayquote}

Another tactic is to use language that sensationalizing in order to convey a sense of urgency or danger to the reader, which will encourage them to share the content further or act on the misinformation in some way:

\begin{displayquote}
``\textit{Look, people do tend to respond to claims that are more sensationalized...And claims that use capital letters more to make themselves sound more urgent...the claims that tend to go viral are the ones that tend to be very sensationalized like `Oh my God guys, we are going to die.'}'' (P5)
\end{displayquote}

% \begin{displayquote}
% “The tone of outrage or anything that stokes emotion, plays on sensations, or even it’s the tone of urgency is going to be spread more” (P15). 
% \end{displayquote}

% \begin{displayquote}
% “A lot of people (are) twisting things that are vague or badly reported onto the most extreme conclusion. And then you also see a lot of religious stuff as well… It's not something I can check but posts like that tend to gain a lot of traction.” (P05)
% \end{displayquote}

% \begin{displayquote}
% “The most important one is sensation building because people always want to click something that's sensational and with more emotions and sensationability. So it's in the first place.” (P16)
% \end{displayquote}

Interviewees pointed out that this tactic isn't always indicative of an urgency to address on its own. 
% While using emotional and sensational hooks can play a key role in appealing to the instinctual or irrational self, P28 believed that ``\textit{they may not be as important as the actual information itself.}'' 
Indeed, P3 pointed out that if it is too obvious that “\textit{someone is trying to instigate something by pitting people against each other},” people might catch on. 
% This is similar to the above quote from P2 about anti-vaccine misinformation, where too much use of sensationalism that doesn't eventually pan out may lead to loss of credibility. 
\vspace{2mm}

\item \textbf{Time sensitivity.}
Most of our participants mentioned instances when a piece of misinformation is conveying a sense of time sensitivity. This was considered important to address because the ways that such messages encourage panic and thus virality. One interviewee talked about cases where the speaker urges the audience to perform certain actions that are claimed to have crucial stakes within a short time-frame:

\begin{displayquote}
“\textit{For instance, this store is closing down or is going bankrupt or being looted or something like that, you need to stock up on this thing. That's very time sensitive because it gives people a time limit and it tells them to go and do something and take some action.}” (P1)
\end{displayquote}

\end{itemize}

\subsubsection{The information context in which the misinformation is interpreted or spread by the public}
%In addition to referencing whether the possible impacts of a piece of misinformation, 
Interviewees also talked about how urgency can be higher or lower based on its context, or how the piece fits into the landscape of other content and claims. When a piece of misinformation taps into a narrative such that people see connections with past happenings, or is responsive to current events and ongoing discussions, it can have outsized impact due to greater receptiveness and contagion within the public---as more people are exposed to the claim as it travels, the more potential for harm. In addition, misinformation can fill a void where there is currently a lack of public information or significant confusion.

\begin{itemize}

\item \textbf{Part of a larger narrative or body of misinformation.} 
While millions of fake or misleading claims circulate around the internet every day, many of our participants pointed out that many individual messages fit into a larger narrative. And together, the larger narrative imposes a more profound impact on our society than any of the individual messages---in other words, misinformation has accumulative effects. Thus, even if an individual claim itself does not directly or immediately lead to physical or societal harm, it may be more important to address if it contributes to a narrative with physically or socially harmful consequences. 
% \begin{displayquote}
% “\textit{When people from one group attack people from another group on the street, that's not usually based off one piece of content. That's based off the long accumulation of events and false beliefs...It's not always just one piece of misinformation...we try to judge every piece of misinformation based on how dangerous the narrative it's playing into is...}'' (P15)
% \end{displayquote}
For instance, when a piece of content is put into a larger narrative such as about anti-vaccination, even an apparently innocent joke could become harmful:

\begin{displayquote}
“\textit{When you may fit this joke into the bigger narrative...for example, there is a picture of Mike Tyson wearing a black t-shirt, and it has the emblem [of] something related to anti-vaccination. On one side, it's just a doctored picture, but on the other side, you may fit it into the bigger narrative that, look, Mike Tyson, he supports the anti-vaccination movement. Then anti-vaccination ideas get reinforced.}” (P2)
\end{displayquote}

Some interviewees also talked about even broader and more longer-term narratives such as age-old biases or prejudices about certain faiths or communities. 
In addition, if there is still existing tension between different groups, such between the different castes and religions within India, misinformation can be considered more urgent to address:

% Something like religious tensions and existing between communities of different caste or between people from different religions as well. So when such kind of narratives, 
\begin{displayquote}
“\textit{...especially there have been riots in the country in the past between people from different faiths...when such [a] thing happens, it can again increase ongoing real-world violence...a very simple child kidnapping rumor in India has caused multiple deaths.}” (P8)
\end{displayquote}

Most interviewees also brought up the accumulative effects of misinformation as a whole, as the flood of false stories over time can overwhelm our capacity to reason and believe. 
Over time, different unrelated narratives may overlap as people make connections among them; one example is the QAnon conspiracy that incorporates elements of many different conspiracy theories and racist tropes \cite{Zuckerman_2019,Bleakley_2023}. 
%This aligns with prior literature on how any kind of misinformation can be harmful to society in aggregate by slowly chipping away at our collective trust. 
Context, therefore, should have some utility for a fact-checker's prioritization criteria, as misinformation that is widely believed and durable over time may arguably be important to address in spite of its lack of direct or immediate impact; some interviewees expressed concerns about the overwhelming nature of competing information in ways that align with research on social media fatigue \cite{Zheng_Ling_2021}. \vspace{2mm}

\item \textbf{Relevance to current events and people.}
As all but one of our interviewees pointed out, misinformation is seasonal in nature. 
Like news, misinformation (around a certain topic) becomes relevant when something happens to catch people’s attention, and fades out when the public shifts its attention to something else.
Thus, beyond the pressure to produce fact-checks that drive traffic to their site, fact-checkers find it is important to fact-check misinformation that ties into current events because people are more likely to share it:

% \begin{displayquote}
% “If something is in the news right now, people are just a lot more likely to share it, I think, versus if there's... the election comes to mind, or the recall election in California that is taking place soon. I've seen some pieces of misinformation about that. Just with that right around the corner, people are much more interested in it and much more likely to share this sort of stuff.” (P11)
% \end{displayquote}

\begin{displayquote}
“\textit{...if you talk today about COVID-19, it makes sense. But if you talk about HIV or polio or measles, it would not be going as viral because people will not be able to connect with [the] majority of the people. Now COVID-19 is something everyone can connect with.}” (P8)
\end{displayquote}

Typically, when major news breaks, misinformation related to it also becomes viral as readers strive to make sense of the situation; attention to the false stories eventually fades away as the news topic itself recedes from the center of public attention. %We have seen this pattern in Boston marathon bombing, presidential elections, and COVID-19. 
Part of the reasoning behind the boom-bust cycle is the desire for novelty both in the news industry and on the part of people:

\begin{displayquote}
“\textit{Now it's about health misinformation. Maybe in a year, if it's all over, it's going to be about something else. I'm really interested in what is going to replace this health misinformation...Now it's an abundance of health misinformation but it's going to fade out eventually. Everyone is bored by all of this. Not a lot of people are damaged by the vaccine. The fakes about `you are going to die' lose their credibility.}” (P2)
\end{displayquote}

Helping to assist these trends is when the content is shared by someone who is highly influential or trusted; for instance P9 had this to say about influencers on Instagram and TikTok: ``\textit{They can be so powerful, especially on younger people because they grow up using them.}'' \vspace{2mm}

\item \textbf{Lack of accessible public information on a topic.}
Interviewees also talked about how sometimes it would be important to address something because of a lack of public information about that topic in the internet or social media landscape, a kind of `information gap.' When there's lack of information, for instance, about an emerging topic, it can be a more urgent situation both because people may be more willing to share it to ``get the word out'' and also because people may be more vulnerable to believing the misinformation:

% We define information gap as the gap in knowledge between experts and the general public regarding the specific topic associated with a piece of misinformation. Our hypothesis was that when the gap is wide, the audience has more blanks to be filled in regards to the topic related to the misinformation. Since any information can fill these blanks, misinformation is therefore more likely to be believed and spread. However, it turns out from our interviews that the information gap plays a tricky role in misinformation virality. Although many interviewees do see information gap as a primary factor contributing to virality, some other interviewees hold different opinions. For instance, the researcher from Africa Check argues that the lack of knowledge of the general audience is probably the most important reason why a piece of misinformation went viral:

\begin{displayquote}
“\textit{...it's playing on confusion or a lack of information somewhere. So for instance, when there is a new strain of COVID-19 discovered, then the misinformation targets that new strain, because people don't know a lot about it yet...So you might see a message that an expert would know is nonsense just by looking at it. But if it's about a topic that the average person won't know much about, then they're going to share it just in case.}” (P1)
\end{displayquote}

% \begin{displayquote}
% “if the audience knows the correct information, then I suppose the audience just ignores the misinformation or just goes into the comments to prove their opinion. If the person doesn't know anything, then the person is vulnerable.” (P02)
% \end{displayquote}

% \begin{displayquote}
% “Organizations that are good at taking things that have a hint of truth to them, particularly on issues that the average person isn't very well versed on like science. And twisting them to take whatever conclusion that they want to push, tends to be the most effective at spreading misinformation.” (P05)
% \end{displayquote}

% \begin{displayquote}
% “And on the case of pandemic, when everything is so not understandable, and even scientists doesn't know at the end, what is going on. Of course, people want answers to questions. And so they decide to believe in something as they decide to choose some answers for them.” (P16)
% \end{displayquote}

In addition to a lack of information, there are also topics that are more inaccessible for laypeople, such as areas related to science. In this case, there is public information but it requires some interpretation by someone with expertise to properly understand:

\begin{displayquote}
“\textit{Just one of the most common reasons pieces of misinformation have gone viral that I've seen is people who don't have the level of expertise necessary to interpret pieces of information misinterpreting that, and misapplying their understanding of that.}” (P11)
\end{displayquote}

% However, when asked to rank the factors that contribute to virality, a fact-checker from Factly told us that the information gap won’t come to the top of his list (“Similarly, difficulty in validation or information gap also doesn't come up.” – P08). 

% \begin{displayquote}
% “Yeah. And I can't think of any way the information gap can influence the spread of misinformation, so I would write it as the last one.” (P17)
% \end{displayquote}

% \begin{displayquote}
% “... a lot of misinformation is not really about the information gap but it's about perception.” (P03)
% \end{displayquote}

However, not every interviewee found this situation to be a criteria for importance.
We noticed that only fact-checkers from regions where science- and health-related misinformation was prevalent ranked the information gap highly, whereas people from places where political, religious, and social conflicts were the main issues deemed it as unimportant. The offline or local contexts, in other words, mattered.

\end{itemize}

% One possible explanation is that the information gap only matters when the misinformation has not been used as propaganda, and the general audience does not have a strong opinion on the related topic. When it comes to science or health issues, most people do not have strong opinions or partisanship. Thus the extent of the audience’s knowledge in the related fields significantly influences how the public would receive the misinformation. However, when it comes to social and religious issues, people have already invested strong feelings and opinions, and they generally do not care whether the information is accurate. Spreading misinformation has become a way for them to propagate ideologies and beliefs. This explanation is supported by a fact-checker from India:

% \begin{displayquote}
% “...depending on the context, information gap actually plays a huge role, especially in terms of vaccines as well… For example, a simple data activate is the percentage of vaccinated people is a lot more in urban areas than rural areas. That clearly indicates information gap (impacts) the narratives that were shared on vaccines…” (P08)
% \end{displayquote}

\subsubsection{Affinity of misinformation for certain communities, cultures, and countries}

In fact, the consideration of offline context raised a final theme by interviewees: the recognition that misinformation has different impacts in different communities and areas with different norms and beliefs. There is, in other words, an affinity of certain kinds of misinformation for specific contexts, where such stories are more believable or more likely to go viral,  making them more important to address. 

For instance, health related misinformation is prevalent in the U.S., and vaccine hesitancy has been a long-standing problem. 
% As one of our interviewees pointed out:
% \begin{displayquote}
% “The United States has had a significant anti-vaccine community even before COVID. Anti-vaxxers have been around for a while now. So I would say that anti-vaxxers to an extent got a little bit of publicity in the last 15 months or so.” (P03)
% \end{displayquote}
However, that isn’t the case for India, where vaccine hesitancy is not as common.
% In India, vaccine hesitancy isn’t a huge issue. 
One participant (P4) mentioned that in India, there is a shortage of vaccine supply compared to the demand, and most media convey that people want to get vaccinated.
% \begin{displayquote}
% “India actually has a vaccine shortage, a shortage of supply. I don't think there's a demand shortage. And increasingly, anywhere, all the media images I've seen show that people want to get vaccinated.” (P04)
% \end{displayquote}
For countries like India and South Africa, misinformation and hate speech relating to social and religious discrimination has been a much more prominent issue.
Understanding the affinity of misinformation requires a clear understanding of the local and cultural contexts on the part of fact-checkers to know which misinformation will have disparate impacts at a local level.

\begin{itemize}

\item \textbf{Misinformation that attacks marginalized groups.} 
Almost all interviewees spoke about how misinformation can be deployed to attack specific communities where there are existing local prejudices. Some highlighted misinformation targeting specific minority religious groups or marginalized communities such as foreigners or migrant workers as the most prevalent. For instance, one interviewee said, ``\textit{I think the biggest threat in India is that 80\% of all the misinfo is targeted towards one particular community, that's Muslims.}” 
%P3 is the source of the quote above but feels too possible to de-identify
Interviewees also spoke about the use of tactics to paint the targeted group as immoral or to dehumanize them:
% \begin{displayquote}
% “\textit{I would say communities are the most targeted. There are so many communal tensions in many parts of Africa, or even against foreigners and migrant workers...}.” (P15). 
% \end{displayquote}

\begin{displayquote}
“\textit{Often it'd be a claim about some horrible thing that the other group would have done---some false claim. It can often be something about how they harm women, how they harm children...accusations of rape---things that are really meant to stoke really strong emotions, and that we see pretty often.}” (P15)
\end{displayquote}

This kind of misinformation can exacerbate existing tensions, play into long-running narratives, and spill over into violence---in short, add to the other aspects already discussed. However, because the targeted group is already marginalized, the importance of addressing this particular kind of misinformation can be heightened for fact-checkers. \vspace{2mm}

\item \textbf{Misinformation targeted to susceptible groups.}
On the flip side of misinformation that targets specific groups is when a group is targeted to be the recipients of misinformation. When these groups are targeted---a concern for most of our interviewees---the goal may be to trap them in financial scams, sell them specific products, or otherwise manipulate them into some action desired by the misinformation poster.
Targeted groups for these purposes include people who may be more susceptible because they have lower media literacy or a lack of resources or motivation in seeking credible information. Thus, members of these groups will not investigate the source or even expect a good source. Other times, susceptible groups are those who have greater fear of the unknown due to being more sheltered or who are generally vulnerable members of society. These were deemed by one interviewee to encompass the following groups:
% \begin{displayquote}
% “If you have low media literacy, you tend not to check the sources, right? The source is not important. So in order to target them, you don't need to have good sources, or not having sources at all. And so a meme is perfect because it doesn't have any source, it's just a claim.” (P13). 
% \end{displayquote}
% But in general, our participants believe that vulnerable groups are most impacted by misinformation.  

\begin{displayquote}
“\textit{Unemployed people, elder[ly] people, more generally people who live in fear of their security. So mostly people who live outside of inner cities, people living in the suburbs, in the countryside...}” (P21) \vspace{2mm}
\end{displayquote} 
% \begin{displayquote}
% “The retired, the elders, they tend to have different interests than the younger, the youth. Misinformation is usually affecting you if it’s something you care about…If it's elder, if you're talking about health, about death or life, then it's affecting them the most, because that's what they care about the most. But if you're talking about the economy, and talking about your job opportunities, then probably the age of the twenties to forties, that will affect. So really depending on the ages, the targeted audience is different.” (P12)
% \end{displayquote}

% \textbf{Impact of misinfo in different regions.} In addition to the threshold of virality, we found that different countries might be susceptible to different types of misinformation. For instance, health related misinformation is prevalent in the US, and vaccine hesitancy has been a long-standing problem.

\item \textbf{Misinformation aligning with existing cultural or political biases.}
Finally, some interviewees talked about the believability of certain claims when it conforms with existing biases. For instance, one fact-checker felt that on the topic of politics, biases about one's opposing side may make many claims seem plausible despite not having evidence:

\begin{displayquote}
“\textit{Sometimes it's utterly believable such as when this happens a lot. The opposition party leader in India, Rahul Gandhi, is portrayed as having said a lot of things that he would not have said. Especially if I'm not positively inclined towards Rahul Gandhi, I will believe pretty much anything that I see so it's highly believable.}” (P3)
\end{displayquote}
%(P3)
% \paragraph{People promote misinfo for personal interests}
% Misinformation can often be used as a tool. It can be wielded to propagate certain beliefs or advance personal interests. 
Misinformation posters and spreaders, some interviewees speculated, also may have some self-interest driving their actions. People might fabricate misinformation to support a cause, politicians might amplify misinformation for their own political gains, and individuals might even spread misinformation supporting a certain belief as a means to gain community recognition. 
\vspace{2mm}

\item \textbf{National differences in preferred media formats.} %We found that different countries might be susceptible to different types of misinformation. 
We also noticed a difference in how fact-checkers from different countries talked about what characteristics help cause a piece of content to go viral. According to a U.S. fact-checker (P11), the kinds of media formats that are most likely to go viral are short text and images. This is because they are convenient to read and can easily catch people’s attention.
% \begin{displayquote}
% “Posts that will stoke communal tensions and possibly lead to physical harm on the streets is something we are very aware of. It's a real thing in Africa where if a piece of misinformation about a certain group of people, especially a minority group goes viral, then there can be real physical repercussions during riots.” (P15)
% \end{displayquote}
% Another interesting insight we obtained from these fact-checkers is that the impact of media format on the virality of a message varies by country.
% Contrarily, a long video requires a lot more patience to watch it from beginning to end, is less likely for people to memorize, and is thus less likely to go viral.
% \begin{displayquote}
% “You're much more likely to see a short Facebook post go viral than you would see a 10-minute long video on Facebook. So, just how that media is formatted. I think people tend to respond much more quick, easy to digest stuff, and are more likely to share that than they were a longer video or slideshow.” (P11)
% \end{displayquote}
However, some of the fact-checkers from India offered a different opinion. From their point of view, videos are usually more influential and provocative than images or text, and therefore, the most effective media format to spread misinformation. 
One fact-checker shared his thoughts on why video is such a popular media format in India:

\begin{displayquote}
“\textit{India has a lot of people, and Indians have a lot of time, and they need to keep themselves occupied. Many of them also are unemployed right now. The economy isn't doing too well, and a great way to keep yourself occupied is consuming any and all content you can find on the Internet.}” (P4)
\end{displayquote}

Fact-checkers are sensitive to issues that are grounded in the communities and language that misinformation operates, thus highlighting how local factors seem to contribute to the spreading of rumors in different ways. As a result, a piece of misinformation designed to go viral in the U.S. may not perform well at all in a different country such as India. %This means that any tools and processes for prioritization must take local considerations into account instead of a one-size-fits-all approach.

\end{itemize}

\subsubsection{Summary}
In sum, multiple considerations inform the approach that fact-checkers take when it comes to prioritizing which claims to fact-check. Consideration around harm appears to be key. Fact-checkers try to assess the possible kinds of harmful impacts of misinformation, alongside the different characteristics of the content itself that seem to increase urgency, like emotional language. Fact-checkers also take into account the larger contexts that surround misinformation: not just within active internet narratives or current events, but also how specific misinformation can affect local communities and countries. 

\subsection {RQ2: How do fact-checkers decide what to fact-check and what tools could improve their processes of prioritization?}

%\subsection{RQ1: What process do fact-checkers take when deciding what to fact-check?}
Having understood what aspects of misinformation create a sense of urgency or importance for fact-checkers, we now inquire more about how they decide what pieces of information to fact-check and whether tools could improve their decision-making processes. We present the following findings about their existing processes.

\subsubsection{Fact-checking has limited capacity.} We asked our interviewees whether they were able to fact-check and publish everything they thought was important, and the answer was overwhelmingly `no.' Fact-checking organizations have limited resources and capacity, particularly if they are a small team. 
Some claims and conspiracy theories are inherently not fact-checkable and are avoided as a time sink.
Chasing down evidence or getting a comment from a primary source can be a time-consuming task and in some cases futile:

\begin{displayquote}
``\textit{We, as a Ukrainian small organization, are not as strong an organization to ask people from [a large international corporation], `Please give us a comment.' I wanted to, I texted them, but all this process of big companies, you text, ask their press office to give a reply, but no one gives you a reply}.'' (P2)
\end{displayquote}

%Yet there is an endless stream of misinformation floating around the internet. 
Fact-checkers face competition from other fact-checkers in some countries, as well as time pressure to fact-check quickly in order to have the most impact. Delivering fact-checks to the right audience at the right time is as important and even as challenging as fact-checking itself:

\begin{displayquote}
“\textit{Within last week, we felt really pressured to fact-check stuff very quickly because the protests were moving at a very fast pace. If we published stuff related to last week this week, it wouldn't be timely anymore.}” (P5)
\end{displayquote}

In light of limited resources, all interviewees acknowledged the necessity of making decisions on prioritizing certain claims to fact-check given competing pressures and current events, even if it meant disregarding other instances of noticed misinformation.

% \begin{displayquote}
% “That also depends on the team availability. We should understand that there are always fakes, but the human resources are limited. People cannot run and follow these fakes to debunk them all the time.” (P02)
% \end{displayquote}

% \begin{displayquote}
% “We've really not been able to arrive at a conclusion so far. Only one fact-checker out of about 15 odd IFCN-recognized fact-checkers in India did that fact-check. 14 others haven't yet been able to arrive at a conclusion, including us.” (P04)
% \end{displayquote}

% \begin{displayquote}
% “But if it is about some other conspiracy theory about some personality or some other sports person, which may not be, where we do not have any authentic evidence as such. So such kind of theories cannot be fact checked” (P08)
% \end{displayquote}

% \begin{displayquote}
% “So especially those time sensitive claims, you want to publish something that is short and very simple to both read and share on because that's the only way it's going to spread before the misinformation.” (P01)
% \end{displayquote}

%As demonstrated by the quotes, fact-checking is urgent and time-pressured. Therefore, it is important to triage misinformation and select out the most harmful ones effectively and efficiently. To better understand the varying harms of misinformation, we consulted fact-checkers on how they triage misinformation in their daily fact-checking.

\subsubsection{Multiple considerations go into decision-making regarding prioritization.} 
 
According to our participants, there were three factors that most agree are important when deciding which piece of claims to fact-check first. 
The first, which is the main focus of this work, relate to aspects of urgency and importance of the claim, which interviewees described as related to harm, virality, and impact. 
However, interviewees also mentioned two other factors that were also important to their decision-making: limited resource allocation, as experienced around the scope of the claim, and strategic considerations due to the interests of other stakeholders.

\begin{itemize}
\item \textbf{Urgency of the Claim}:
As revealed in Section \ref{sec:urgency}, interviewees validated our hunch that fact-checkers do prioritize what is urgent to fact-check based on a harms analysis, which typically incorporates the virality or reach of the claim, the possibility for negative consequences (particularly physical harm), and the potential impact the misinformation will have on society:

%The first important element would be the virality of a piece of content: how far it has spread, and how fast it is spreading. The second major factor is the potential for imminent real-world harm: fact-checkers evaluate the toxicity of the content and look for messages that might sway people into taking actions that may harm others or themselves. 

% \begin{displayquote}
% “If it's something that has the potential to spread beyond what it already has, so if it seems like it's particularly viral, if it's about a topic that's in the news right now that people are going to be interested in and that hasn't been widely covered yet, if it has the possibility for real-world harm.”
% \end{displayquote}

\begin{displayquote}
“\textit{[We triage misinformation] by how widespread the claim is. So if lots of people are going to act on it, even a small harm can get amplified; and also by how harmful it is for each individual person to act on it. So a claim that tells someone to take an untested medical remedy might be very harmful to everyone who tries it because they take something poisonous.}” (P1)
\end{displayquote}

%In the next subsection \ref{sec:urgency}, we dive into more detail regarding the specific characteristics of misinformation that make them more urgent or important to address. 
Our interviewees also reported facing difficulties in estimating the potential harm and impact of a piece of misinformation; for example certain app designs, such as E2EE messaging apps, create barriers for estimating virality. \vspace{2mm}

% \begin{displayquote}
% “Virality or the reach of a post, definitely is a factor. We then also look at it in terms of is there something else to fact-check that needs to be answered first? Or what is the potential harm that this could cause?” (P04)
% \end{displayquote}

\item \textbf{Resource Allocation and Claim Scope}: 
Interviewees also described taking into account their own limited resources and capabilities, as well as the role of fact-checking organizations and their relative strengths compared to other kinds of news or information sources. In terms of scope, fact-checkers usually prefer prioritizing claims that can be clearly refuted in a timely manner with real evidence.
% \begin{displayquote}
% ``\textit{There'd been times when my fact-checks had been parked because, I'll try fact-check something, there is no real answer either way and my editor then says, `Well this isn't adding anything new to the conversation so let's rather not publish.}''' (P06)
% \end{displayquote}
As a result, claims that are too hard to validate such as  rumors, outlandish conspiracy theories, opinions, predictions about the future, or vague statements are all considered out of scope:

% \begin{displayquote}
% “\textit{We don't fact check things in the future, for example, two days later there's an earthquake...Or if it's national security matters, sometimes we are not able to fact check...For example, once we had a user claim...the Defense Department had a missile park in Tainan...Is that true or false? ...even if we know, we cannot publish that to the public unless the Defense Department did announce it themselves. Yeah, those kind of criteria will also change the priority of what we fact-check.}” (P12, Taiwan)
% \end{displayquote}

\begin{displayquote}
``\textit{...If someone says, `I think that Biden is a better president than Trump', we can't fact-check that... If someone says, `By 2050 the ice capsule will melt', we can't check that. And also we can't really check very vague things, so if someone just says like, `A lot of people think, I don't know, that guns should be banned', you can't fact-check that because what is a lot of people? You need to acknowledge your own limitations, so what we really look for when we look for things to fact-check is something very specific...}'' (P5)
\end{displayquote}

Beyond claims that are out-of-scope, if a claim takes too long to fact-check or requires too many resources, fact-checkers tend to move on to claims that are easier to fact-check and come back to the `hard' claims only after the other ones have been handled. \vspace{2mm}

\item \textbf{Interests of Different Stakeholders}: 
 Finally, fact-checkers often have to weigh the (potentially competing) interests of different stakeholders before deciding on which claims to fact-check, e.g., their own organizational interests and financial incentives, needs of media platforms with whom they are collaborating, the public's interest in the topic or its newsworthiness. For instance, several interviewees mentioned partnerships their organization had with social media platforms (e.g., Facebook, WhatsApp, TikTok), where those platforms had content that they wanted to prioritize for fact-checking.
%  \begin{displayquote}
% "\textit{...we're third party fact-checkers to social media platforms like Facebook, WhatsApp, ShareChat, TikTok outside of India. So when we are reviewing content on these platforms, many of them have a queue system where we get to review content.}" (P04)
% \end{displayquote}
Some interviewees working with certain platforms talked about a pressure to write as many fact-check pieces as possible as they were paid according to how many they complete. In terms of other financial incentives, some interviewees also talked about a pressure to fact-check items that are already being widely discussed, which would then drive more traffic to the fact-check site:

\begin{displayquote}
``\textit{I think the editors would love it if we were constantly fact-checking things that were front page news, but it also means that there's a bit of a pressure to focus on things that are maybe already getting widely spread and widely shared and talked about.}'' (P1)
\end{displayquote}

% \begin{displayquote}
% "I suppose that organizations which work with Facebook, they may have financial incentives to write the debunkings. Facebook pays for each piece of the material. Some of the organizations, I think including mine, has the motivation to write as more as possible pieces to receive more money." (P02)
% \end{displayquote}

% \begin{displayquote}
% "That also depends on the team availability. We should also understand that there are always fakes, but the human resources are limited. People cannot run and follow these fakes to debunk them all the time." (P02)
% \end{displayquote}

\end{itemize}

\subsubsection{No established systematic approach towards harm assessment or claim prioritization.} 
Finally, we found that across our interviewees, few described any kind of established or systematic method they or their organization employed to assess the potential harm associated with a piece of misinformation or otherwise prioritize the most harmful claims. Current practices of claim selection tend to be ad hoc and time-consuming.
It is often also a collaborative decision-making process involving the entire team of fact-checkers, editors, and sometimes media partners.

% \begin{displayquote}
% "If I have too much on my plate, I'll either ask another researcher to deal with it or I might decide it's not worth dealing with it." (P06)
% \end{displayquote}

% \begin{displayquote}
% "So we have a team of people. So basically, probably our day starts with looking for claims across all social media platforms. We have active WhatsApp people. A number of people tag us on Facebook, Twitter. Sometimes people directly email us. So over the day, we usually receive a lot of request or content to fact check. So each team member has specific roles. So most of them go through in the first one, they're on office, they go through all the social media claims or some WhatsApp requests that we receive. And once it's done, they will sit with me to decide on what can be fact checked, what cannot be fact checked" (P08)
% \end{displayquote}

\begin{displayquote}
``\textit{And when we find something valuable, something that's spread really quickly. And that reflect a topic that is important now. We discussed it between our team members. We have different channels where we discuss the things, like in Messenger or in our group on Facebook. And then we decide, should we take it or not?}'' (P16)
\end{displayquote}

In some cases, interviewees described disagreements between members of the team where they would debate what to cover. In other cases, individuals got to choose what they wanted to fact-check but might get suggestions from other team members or get advice from a more senior member:

% \begin{displayquote}
% ``\textit{So there is often this debate sometimes within the team itself. I come with certain views and some other person could usually come up with some other view...When we see the harmfulness of the narrative, there could be sometimes a discussion or is there anything that we can go ahead and fact check it?}'' (P08)
% \end{displayquote}

\begin{displayquote}
``\textit{So first, we have a stage, that's called the media scan, where we choose our topics and...[the] whole team can give us suggestions what to choose... Typically, [one] chooses the subject he wants to write about, but sometimes he's advised by, for example, me or some other editor regarding, for example, whether we fact-check this article...from the view of [the] fact-checker, they would most likely ask someone more senior for their opinion, which to fact-check.}'' (P17)
\end{displayquote}

% \begin{displayquote}
% "But usually, it will be politician statements. If there is a big speech that either the head of a political or the president is giving, we might listen in to see if there is anything to fact-check there. Sometimes we listen in to parliament and see if the people in parliament are making claims so we can fact-check. We really get our claims from a variety of sources." (P06)
% \end{displayquote}

%\subsection{RQ3: What tools would be helpful to fact-checkers with regard to prioritization?}
%For our final area of inquiry, we ask questions about what kinds of tools could help fact-checkers in their effort to prioritize inaccurate or contested claims.

\subsubsection{Tools evaluating potential harmful impacts of misinformation}
Our participants mentioned the usefulness of evaluation tools that could help them understand the potential impact of a piece of misinformation. Interviewees hoped, for example, that automated semantic analysis or detecting keywords that are related to hate speech or provocations of violence could help them assess the degree of potential harmfulness. Some interviewees also told us that auto-extracting the topics covered in a piece of content and evaluating their relevancy to local events would significantly speed up their process to select more urgent claims to fact-check: 
% A more detailed discussion on tools accounting for local knowledge would be conducted in the next subsection.

\begin{displayquote}
``\textit{I would like it to take care of the urgency of issues in society. For instance, I would like such a system to weigh COVID-19 misinformation and political misinformation more heavily. And that's considering my own environment. So maybe in the U.S., it could be something else that is more urgent.}'' (P19)
\end{displayquote}

In addition, a few participants pointed out that such a computational tool should not only display a score or level of harm, but also show an analysis on why it is harmful:

\begin{displayquote}
``\textit{I think it would be useful to have almost at a glance, not just an understanding of what is the most harmful misinformation, but also why... It's good to be able to look at something and see, oh, this is going to be harmful because it's widely shared, and this is going to be harmful because it has a very direct impact, that helps with the prioritization.}" (P1)
\end{displayquote}

%\subsubsection{Summary}

%[some kind of summary/takeaways here?]
Thus, fact-checkers desired to more quickly and fully understand a range of potential harmful impacts based on differing aspects, whether direct threats of physical violence to a piece of misinformation's affinity for the local context.

\subsubsection{Better tools to track and measure current and potential spread of misinformation}

The easiest and most direct way to estimate a claim's virality %would be measuring
is to measure the current spread of the claim. Most interviewees reported that they found tools that show the number of viewers, the share speed, 
%of being shared, 
and the reach and shares of a claim across demographics useful. And for fact-checkers in international organizations, they pointed out that there are yet no tools (though demanded) that could display a global view of misinformation spread.
% \begin{displayquote}
% "It's more out of interest in seeing how trends spread globally, because we do see so many claims that are international. We do translate a lot of fact checks from one language to another. We do adapt a lot of our fact checks with the same proof but different claims because it's slightly changed in different countries. I mean, that's what's interesting being an international fact checking organization is how much things are repurposed from one region to the other, and how claims spread." (P15)
% \end{displayquote}

Fact-checkers also pointed that detecting the variations and adaptations of a claim %to 
across different platforms or media formats could also be a strong indicator of how popular a claim is. If a software can tell that the same message has been adapted to both text and video, and it is spreading in Twitter, Facebook, and other media platforms, then the fact-checkers could be fairly confident that such a claim is truly trending at the moment:

\begin{displayquote}
``\textit{Maybe something that could link narratives, like this story looks like this on Twitter; it has also spread on Telegram and it looks like this. So to be able to connect similar stories throughout the internet and have it visualize for every piece of misinformation I should fact-check. That would really help with monitoring things that has gone viral.}'' (P14)
\end{displayquote}

In addition to measuring the current spread of a piece of misinformation, %another important part in measuring  
a robust metric around virality would also aim to predict the spread of misinformation. Detecting and flagging claims with the potential to go viral, and before the public has been been widely exposed, would greatly assist fact-checking organizations.

Predicting the potential spread of misinformation requires an estimation regarding the factors that make up virality, a topic around which fact-checkers had varying thoughts. For instance, some fact-checkers believed that tools accounting for the popularity of the sender would give them a good sense of how well the claim would spread; others thought evaluating the believability of a claim would be a good signal. %More importantly, 
The majority of our interviewees desired computational tools that could flag a content by its sentiment level or the use of specific words.
% \begin{displayquote}
% "If there was a tool that could flag up some things that are more emotional and more likely to be shared, maybe that would be interesting." (P15)
% \end{displayquote}
% \begin{displayquote}
% "\textit{A tool that can target specific words can give me some insights into what is being discussed and what tends to go viral. I would put the keyword in this tool, and that tool would give me all the results when exactly it appeared on social media. And then I can choose by date to see which came first, which came second, and so on. If I didn't have this tool, I would have to sift through tons and tons of information in order to come across them.}" (P6)
% \end{displayquote}
Finally, because misinformation has its own life cycle, understanding where a claim is within that cycle would help the fact-checkers better evaluate the future spread of the claim:

\begin{displayquote}
``\textit{It would have to measure not only the virality, but at which point are we in the viralization? So if it's starting to viralize, if it's already past its peak point then it's coming down, or if it was already very viral but now isn't anymore.}'' (P10)
\end{displayquote}

\subsubsection{Tools that better address local context}

Many participants expressed concerns about using automated tools for selecting and evaluating claims since most of the contemporary tools do not take local knowledge into account. Fact-checking organizations usually have local experts working on the ground to provide context to understand the misinformation spreading locally. An interviewee from France told us that a computational tool can only provide substantial value if its analysis is comparable to that of a local expert:

\begin{displayquote}
``\textit{For the claim itself, I think it really does take [a] human with local knowledge to understand how important it is. Sometimes, like I said with our fact-checkers on the ground, they might understand the importance of something more than...even if it's in English and I can read it and understand it, they might understand more of the context and understand why it's important than some other people in another country. I'm not sure there's a tool that could really do that better than local knowledge.}'' (P15)
\end{displayquote}

Participants also pointed out that it would be helpful to have tools that reveal the differences in a claim's reach and shares among different demographics, as well as any targeted groups/communities if relevant. Thus, if a rumor reached a particular vulnerable group, the tool could help them know that it should be prioritized. 

% \begin{displayquote}
% "In terms of outputs, maybe when you use it you can immediately see the metrics of where it's been shared, and the kind of demographics it's been reaching." (P09)
% \end{displayquote}

\begin{displayquote}
``\textit{It would be a tool that says, for example, this category of persons would be more vulnerable to questions about migration or health or safety. If the misinformation reached the category of people, who are more vulnerable to that kind of topic, then we could prioritize it.}'' (P7)
\end{displayquote}
But if a false claim did not have much reach, perhaps the fact-checking group would hold off; fact-checkers explained that giving more visibility to the misinformation when it has low reach was counterproductive.
% \begin{displayquote}
% "A scenario will be, for example, during an electoral campaign, where a politician claims something which is wrong, and we would have to make a decision on whether or not we fact-check it. If the misinformation did not reach the target of the politician, and we know that if we fact-check this misinformation, we will give more visibility to the misinformation. So we would like to know when fact-checking is efficient or counterproductive." (P07)
% \end{displayquote}

% \begin{displayquote}
% "I would like it to be sensitive to the kind of environment that I work in. For instance, I was in Europe before. So the issue in Europe is not dearth of information. People have enough information in Europe. It's just for them to determine which one to believe, or not. Over here, in Nigeria, and most parts of Africa, you realize that there's not much information. So I would like it to factor that into the effects." (P19)
% \end{displayquote}

\subsubsection{Tools for collective decision-making on prioritization}

Furthermore, several fact-checkers pointed out the need for tools that facilitate transparent and democratic team decision-making. Areas that a tool could assist included broadening the possibilities for participants in decision-making in ways that permitted a discussion of reasons behind certain decisions:

\begin{displayquote}
``\textit{I think any tool should allow us to make decisions in a more transparent way. And in that sense, if more people are involved, everyone can see what the process is, why decisions are made in a certain way, and everyone can contribute to the discussion. That would be much better than having one person centralize the decision-making.}'' (P10)
\end{displayquote}
In addition, one interviewee cited some concern around the potential subjectivity of decisions, especially if the power of those decisions are concentrated in one individual within a team or organization:
\begin{displayquote}
``\textit{I read about the %David Murray's
[...] research 
%that he 
[where someone] asked the editor how he determines which news reports go out, and which ones don't go out. And eventually it was determined that the decision-making for this editor is usually subjective, not necessarily guided by any journalistic approach. So the number one thing I would like to see in this tool is a cleansing of any form of bias that could come into fact-checking and editing. For me, as an editor, I would like to see that.}'' (P19)
\end{displayquote}

\subsubsection{Summary.}
In sum, we validated that fact-checkers do make decisions regularly about prioritization. These decisions often reference signals or concepts that relate in some way to harm.
%Signals that fact-checkers consider include virality or spread, which affects the number of people who are exposed to the claim and may be harmed, and the possible negative consequences of believing certain claims, which can directly harm individuals.
% These concepts form the basis of our exploration into a harms model for misinformation.
We also note that fact-checkers have additional organizational and strategic considerations not covered by our work to develop a harms model of prioritization. Finally, we find that fact-checkers do not have an established process and desire additional support, suggesting that tools to support structured consideration of harm could be valuable.

\section{Misinformation Harms Framework for Prioritization}
\label{sec:framework-dev}

% Intro’ing the idea of a harms framework as encompassing what fact-checkers are talking about (“our approach” from whitepaper)

Based on our interview results, we learn that gaining precision about the relative harmfulness of a piece of misinformation can improve the fact-checking process (and experience) of prioritization, by making it less dependent on ad hoc judgment. In addition, misinformation harm assessment can help by more clearly articulating the kinds of impact that fact-checkers seek through their work, and provide opportunities through which their work may be strategically made easier or evaluated through supporting tools.

% For example, organizations that seek to measure the volume of misinformation spreading on a platform or determine the impact of an intervention to reduce misinformation spread may be more interested to focus their measurement on harmful misinformation that affects larger populations.  Other organizations may seek to improve misinformation around a specific topic, even though it may not have broad reach in certain platforms or language communities. 

A definition and evaluation framework expressed as an urgency checklist or questionnaire around misinformation as a harm is one mechanism that both helps fact-checkers prioritize their efforts while also laying the foundation upon which impact could be assessed.  
This work can also serve as annotation, training, or evaluation guidelines for future work in automated or crowd-assisted harm assessment support.
We present the results of our effort to develop a more structured and systematic misinformation harms framework below.

\subsection{Method}
 %    - ("looking to other approaches" in the whitepaper)
%Both misinformation and harm are complex concepts to address. 
To lay the groundwork for this framework, we looked first to researchers who in recent years have developed taxonomies and definitions regarding harm and misinformation; these examples supported our effort to identify harmful misinformation and to establish prioritization.  Then, using an iterative approach, we created a questionnaire that aimed to isolate distinct dimensions of misinformation as harm that could signal a degree of urgency. We incorporated thematic concepts for evaluating harm that we discovered in our review of existing literature, as well as insights derived from past engagement with those working in fact-checking. Through internal sessions and in workshops with other experts, we further distilled these dimensions as we refined the questionnaire. Our final steps included incorporating information from our interviewees into the questionnaire, and then gaining feedback from them and other fact-checkers \add{(\autoref{fig:process})}.
\begin{table}[]
\small
    \centering
    \begin{tabular}{|p{18mm} | p{21mm} | p{23mm} | p{19mm} | p{16mm} | p{22mm}|}
        \hline
          & 
          \textbf{Fragmentation} &
          \textbf{Actionability} & \textbf{Believability} &
           \textbf{Likelihood of Spread} & \textbf{Exploitativeness}\\  \hline
         \textbf{Agrafiotis et al. \cite{agrafiotis_cyber_2016, agrafiotis_taxonomy_2018}} &  Societal harm; Indirect harm; Long-term harm; Community- and Nation-level impacts & Physical harm; Direct harm; Short-term harm & Community-level impacts &  & Psychological and Emotional harm (fear); Group identity  \\ \hline 
         \textbf{FullFact  \cite{fullfact_framework_2020_1}}  & Polarization & Gravity (or severity of harm); Polarization & Novelty  & Scale & Demographics; Novelty   \\ \hline
         % \textbf{McCright et al. \cite{mccright2017combatting}} & & & & & \\  \hline 
         \textbf{Scheuerman et al. \cite{scheuerman_framework_2021}}& Coordinating Scams and Political Attacks & Coordinated Attacks (two categories); Targeted Attacks; Violence; Hate Speech & & & Direct Harm to Children \\  \hline 
 \textbf{Tran et al. \cite{tran_investigation_2020}}& Trust harm; Connection harm & Life harm; Injury harm; Discrimination harm & Confusion harm & Reputation harm; Likelihood of harm (general) & Emotion harm; Isolation harm \\  \hline 
        \textbf{Wardle \& Derakhshan \cite{wardle_fake_2017, Wardle_Derakhshan_2017}} & & (Hate speech) & Imposter content; Fabricated content &  & \\ 

         \hline
    \end{tabular}
    \caption{Aspects of existing frameworks and taxonomies that intersect with our proposed framework of five urgency dimensions of misinformation harms.}
    \label{tab:frameworks}
\end{table}
\subsubsection{Building from existing taxonomies}
Several types of work we referenced include general harm-oriented taxonomies, misinformation-oriented taxonomies, and urgency-related taxonomies. The fuller literature review can be found in Section 2.  A few of these works proved instrumental to our thinking as we clarified our own dimensions and are included in \autoref{tab:frameworks}. 

For example, from harm-oriented literature, work from Agrafiotis et al. 2016 and 2018~\cite{agrafiotis_cyber_2016, agrafiotis_taxonomy_2018} defined types such as \textit{physical}, \textit{economic}, and \textit{psychological} harms, and considered the broadness of impact from \textit{individual} to \textit{national} to \textit{social} or \textit{cultural} levels. The papers also considered the dimension of time, or \textit{short-} versus \textit{long-term} harms. These categories of harm informed four out of five of our dimensions and are represented in some part of our framework.

In another example, work led by the non-governmental organization FullFact sought to describe indicators that could help groups determine whether or not urgent coordinated action might be warranted. One of their papers described how aspects of \textit{Scale}, \textit{Demographics}, \textit{Novelty}, \textit{Polarization}, \textit{Gravity} (or severity of harm) might help indicate the urgency around the need for action or support~\cite{fullfact_framework_2020_1}. We used FullFact's framework to confirm various aspects of urgency represented in all five categories of our framework. 

Scheuerman et al. brought together numerous severity-related concerns from a harms-oriented approach based on platform classifications ~\cite{scheuerman_framework_2021}. However, few of the categories intersected directly with the topic of misinformation; the ones that were most related are listed in the table. Overall, the paper also provided helpful insights with regards to the effort around severity-related classification itself.

Last, for a misinformation-oriented approach to taxonomization, Wardle and Derakhshan's categories of \textit{Imposter content} and \textit{Fabricated content} \cite{wardle_fake_2017, Wardle_Derakhshan_2017} were especially noteworthy to us, as they demonstrating clear cases when people's belief and trust are potentially manipulated. \textit{Hate speech} was also noted, though in their categorization it falls under ``mal-information'' or `genuine' but harmful content.   

\begin{table}[]
\small
\begin{tabular}{|l|llll|}
\hline
\textbf{\#} & \textbf{Gender} & \textbf{Country} & \textbf{Role} & \textbf{Format} \\ \hline 
R1          & Male           & Italy         & Editor     & Interview      \\ \hline
R2          & Female         & France          & Editor/Journalist     & Survey   \\ \hline
R3          & Male           & France          & Fact-checker    & Survey    \\ \hline
R4          & Male           & South Africa  & Fact-checker     & Survey      \\ \hline 
\end{tabular}
\caption{Demographic details of reviewers providing questionnaire feedback.}
\label{table:reviewers}
\end{table}

%When reviewing this literature, we found at a high level, two types of high-characteristics regarding misinformation and harm.  
When reviewing this literature, we found that efforts sought to distinguish or characterize misinformation and harm along two distinct lines that do not overlap.  
First, there is a focus upon defining different general types or \textit{categories} of harm, e.g., physical harm, individual harm.
% For example, categories of harm may be physical versus economic harm, or harm that affects an individual versus society at large. 
At the same time, other efforts capture \textit{variable magnitudes} of harm whose value may vary according to context. 
For example, the reach or `virality' of a piece of misinformation can have a value that is higher or lower depending on factors such as the popularity of the poster. 
When characterizing misinformation as a harm in order to assess urgency, clarifying the difference between these \textit{categorical and variable} characteristics can be useful. It is clear among variable characteristics when a case is more urgent, for example, when something has more virality as opposed to less. However, there may always be disagreements and considerations among which categories of harm matter more: are issues with short-term impacts always more important to address compared to those with long-term ones? Thus, it is easier to define degrees of urgency within a category of harm as compared to across categories.

\subsubsection{Taxonomy and questionnaire development}

 % Using an iterative approach, we created a questionnaire that aimed to clarify distinct dimensions of harmful misinformation around the issue of urgency. 
In addition to reviewing and incorporating thematic concepts for evaluating harm that we discovered in our review of existing taxonomies, we incorporated insights from our own past engagements with fact-checkers. Through informal workshops and conversations with other experts, we also received feedback as we iterated on our taxonomy and set of questions as a team over the course of about a year. 

% we also insights derived from past engagement with those working in fact-checking. Through internal sessions and in workshops with other experts, we further distilled these dimensions as we refined the questionnaire.

As we conducted and analyzed the interviews we conducted with fact-checkers, we began incorporating their themes into the taxonomy. Their considerations regarding the harmfulness of misinformation not only validated our approach, but suggested additional areas for consideration. For instance, the creation of a dimension that we call ``social fragmentation,'' which addresses longer-term community or societal level harms, resulted specifically from their observations addressed in Section 3.3.1 and 3.3.4. 

\subsubsection{Evaluation}
As the framework became more finalized, we turned towards a more formal feedback process. In our final step, we invited four experts in misinformation and fact-checking to work through our questionnaire and share their feedback (\autoref{table:reviewers}). With each expert, we either conducted a 30-minute interview or asked them to complete a survey which guides them through the questionnaire and prompts them to provide feedback on each of the five dimensions. 

All reviewers agreed that our framework matches with their own understanding of misinformation harm and could prove useful in multiple scenarios. 
R4 mentioned that ``\textit{A framework like this would be very useful, and my organization already uses an informal, unwritten version of this kind of framework.}'' 
Similarly, while R3 pointed out that he himself wouldn't need to use the framework as he already had abundant experience in defining harm, he thinks the framework ``\textit{could definitely be useful during teaching sessions or for younger journalists.}'' 

At the same time, reviewers offered insightful suggestions on the questions' coverage, readability, and conciseness, which we incorporated to further refine our framework. 
 For instance, R1 requested greater clarification about narrative on a question within our ``Social Fragmentation'' dimension that originally stated ``\textit{Does the message fit into a larger narrative that has been existing for some time?}'' In response, we modified the question. Using the work of psychologist Jerome Bruner on narrative \cite{Bruner_2010}, the new version reads: ``\textit{Does the message fit into a larger story or argument, for example about how the world works or how people think?}'' with a note explaining its significance: ``\textit{A larger narrative may include stories about communities, race, poltitical parties. A larger narrative crosses platforms and has existed for some time.}'' 
% For instance, multiple reviewers questioned the purpose behind our question ``\textit{Is the message content complicated to understand?}" in our dimension on ``Exploitativeness.'' 
% Hence, we added an annotation for this question, ``\textit{Do you have a hard time reading the content all the way through and explaining it to others?}'' to further clarify its purpose and usage. 

\begin{figure}
    \centering
    \includegraphics[width=1\textwidth]{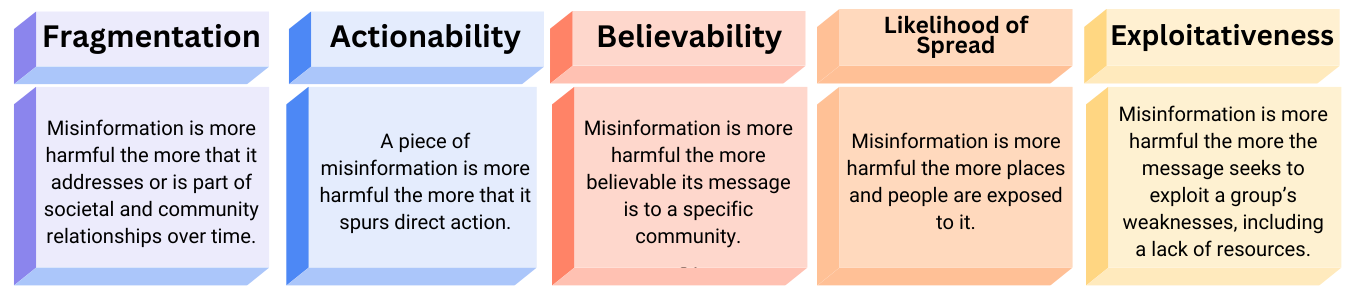}
    \caption{\add{Dimensions of the FABLE Framework of Misinformation Harms.}}
    \label{fig:fable}
\end{figure}

   %  - include table with other frameworks and what they miss/what we use?
%Some examples of work around taxonomies and frameworks that we found inspiring include:

\subsection{\add{FABLE:} A Five-Dimension Urgency Framework for Assessing Misinformation Harms}

We present a framework of five variable ``dimensions of urgency,'' along with  a questionnaire, that can support fact checkers in their efforts to discuss and prioritize in a more strategic way \add{(\autoref{fig:fable})}. These five dimensions can help to clarify urgency to address a piece of misinformation, thereby helping response teams have a clearer understanding of what impacts they may want to achieve. It may be that certain organizations prioritize certain classes of fact-checking content. 
But by providing multiple dimensions, fact-checkers have a view to a holistic approach that reveals what issues might be missed if organizations are always only focused on one or a few dimensions.
% The five dimensions can also be further developed in future research around what is possible to automate in practical interventions.

Our \add{\textbf{FABLE Framework of Misinformation Harms} is a} model that includes five dimensions of urgency when it comes to assessing and prioritizing misinformation as a harm: \add{ \textbf{(social) Fragmentation}, \textbf{Actionability}, 
\textbf{Believability}, 
\textbf{Likelihood of spread},  and \textbf{Exploitativeness}.}
The five dimensions of urgency are each defined through a set of questions, which are incorporated into a single multi-part questionnaire (see Appendix A and online resource\footnote{\url{https://www.artt.cs.washington.edu/analysis-framework-online-misinformation-harm}}); the questions that make up the questionnaire reflect factors that are currently understood to have an impact on misinformation’s magnitude or potential harm. 

% Some aspects of the questions are context-agnostic, such as whether a ``call to action'' takes place in the content. Other questions are both content and context dependent, for example, directly mentioning vulnerable populations that have been the target of misinformation campaigns. 

In the following, we define each dimension according to the relationship between misinformation and harmfulness, which is potential harm or having the capacity to inflict harm. We provide key questions associated with each dimension as well. 
Appendix A has the full list of questions associated with each dimension along with tips for and examples of what to look for.

\subsubsection{Dimension: (Social) Fragmentation}
\begin{displayquote} Definition: A piece of misinformation could have indirect, societal, and accumulative effects.
Therefore, a piece of misinformation is more harmful the more that it addresses or is part of societal and community relationships over time.
\end{displayquote}

This set of questions address the potential of misinformation to affect larger, community-based relationships over time. Issues of peer-to-peer and institutional trust are examples, where a long-term consequence of misinformation is reduced trust in existing institutions and social groups. 
% This category emerged out of our exchanges with fact-checkers and work on a related project on trust. 
Trust and trustworthiness are important issues that must be grappled with by functioning societies and democracies. Indeed, scholars have argued that a certain amount of distrust may be necessary, for example \cite{Hardin_2006}. Hard questions about the appropriate and just functioning of public institutions, the scientific community, or the media are appropriate to ask. However, our framework is focused on the effect of repeated questions that sow mistrust when combined with incorrect or inaccurate information.

Questions include:
\begin{quote}
\begin{tcolorbox}[standard jigsaw, opacityback=0]
\begin{itemize}
\item Does the message fit into a larger story or argument, for example about how the world works or how people think?
\item Does the message question trust in or the functioning of public institutions?
\item Does the message question trust in or the functioning of the scientific community as a whole?
\item Does the message question the functioning of or trust in news sources/ the media in general?
\item Does the message question the trustworthiness of other people in general within a community or society?
\item In a democratic country where there are elections, does the message directly attack the election process?
\end{itemize}
\end{tcolorbox}
\end{quote}

\subsubsection{Dimension: Actionability}
\begin{displayquote} Definition: A piece of content is more harmful the more that it spurs actions that directly cause harm.
Therefore, a piece of misinformation is more harmful the more that it spurs direct action.
\end{displayquote}

Questions tied to this dimension are intended to ascertain whether characteristics or factors related to the message(s) make the content likely to spur directly harmful actions. For example, an explicit “call to action” is a key example of actionability, though there are ways that this call can be obscured. 
We focus on key questions for this dimension on more overt signals regarding direct action or coordination; additional questions listed in Appendix A attempt to capture subtler considerations that may heighten actionability. These include messages that cast aspersions on particular groups, make use of injustice or moral outrage, or are tailored or addressed to communities with a history of violence.

Overall, the \textit{actionability} category emphasizes the potential for physical harm over other types, a characteristic recognized both in economic risk assessments and harm evaluations. The focus on physical harm, when considering actionability, was affirmed in our conversations with fact-checkers. Key questions regarding \textit{actionability} include:

\begin{quote}
\begin{tcolorbox}[standard jigsaw, opacityback=0]
\begin{itemize}
\item Does the message content include an explicit call to action?
\item Does the piece of content incorporate coordination efforts, such as dates/times or other arrangements for follow-up?
\item Does the message provide a name or otherwise any identifying information about an individual, an address, or a place of work in such a way that people might be directly harmed?
\end{itemize}
\end{tcolorbox}
\end{quote}

\subsubsection{Dimension: Believability}
\begin{displayquote} Definition: A piece of misinformation is more harmful the more believable its message is to a specific community.
Related: A piece of content is more effective the more believable its message is to a specific community.
\end{displayquote}

Our believability questions are related to topics where either authoritative consensus is difficult to achieve, or such consensus is affected by the perceptions from a specific community (“in-group”). Answering questions surrounding believability will require at times having a specific community in mind. 

Key questions in this dimension are focused on the inability of readers to easily verify information, whether strong communities or audiences already exist around certain topics, and whether publishers have unclear editorial practices. Other questions take note of the absence of corroborating evidence around certain issues as well as the familiar tone of the messages. The issue of imposter content or accounts are important to this category, as belief is achieved by taking advantage of a community's good will.

Key questions include:
\begin{quote}
\begin{tcolorbox}[standard jigsaw, opacityback=0]
\begin{itemize}
\item Is there a lack of high-quality information that is publicly accessible and is refuting the message’s claim?
\item Does the poster and/or organization/outlet have a noteworthy number of social media/community followers?
\item Is the content published by an organization/outlet with uncertain editorial control (e.g., is not a recognized news publisher)?
\end{itemize}
\end{tcolorbox}
\end{quote}

\subsubsection{Dimension: Likelihood of Spread}
\begin{displayquote} Definition: A piece of harmful content is more harmful the more places it appears on, and the more people who are exposed to it.
Therefore, a piece of misinformation is more harmful the more places and people are exposed to it.
\end{displayquote}

Questions in this dimension try to ascertain whether characteristics or factors related to the message(s) make the content likely to spread or be discovered. It focuses on questions related to magnitude of exposure or potential exposure rather than analyzing the message for its credibility. Misinformation literature often focuses on this vector when thinking about potential impact and, in our interviews, fact-checkers mentioned virality often when considering their own evaluation of a claim’s urgency. 

Our key questions focus on the accounts or persons with histories with large reach or repeated instances of advancing rumors. Other questions go beyond to look at aspects of information context, such as platform design, and characteristics of the content itself, such as direct appeals to the audience or its format (e.g., text versus audio or video). We also included considerations of current events and novel trends, as well as the tone of the message.

Key questions include:
\begin{quote}
\begin{tcolorbox}[standard jigsaw, opacityback=0]
\begin{itemize}
\item Do the people or entities who are spreading the piece of content have a broad reach (size of following on social media, “influencer,” presence on TV or other news media)?
\item Are the people or entities known to be repeat spreaders of questionable information?
\end{itemize}
\end{tcolorbox}
\end{quote}

\subsubsection{Dimension: Exploitativeness}
\begin{displayquote} Definition: A piece of misinformation is more harmful the more the message seeks to exploit human or a group’s weaknesses, including a lack of resources.
\end{displayquote}

These questions addressing \textit{exploitativeness} recognize that factors ranging from emotional manipulation to a lack of available resources can contribute to a group’s vulnerability to misinformation. Harm frameworks that note the vulnerability of groups such as children and the elderly are related. This dimension strives to examine when aspects of the message itself directly engage in the exploitation of human fears and emotions. 

Our key questions highlight common strategies for exploiting particular weaknesses or vulnerable groups. The remaining questions, which can be found in Appendix A, go farther to ask about additional vulnerable groups, such as veterans or conspiracy theorists, or feelings of isolation, powerlessness, or disenfranchisement. We also consider whether the content is in a less popular language, where there are fewer protections and resources for fact-checking, or is spreading in a region where local audiences may be more vulnerable.

Key questions for this dimension include:
\begin{quote}
\begin{tcolorbox}[standard jigsaw, opacityback=0]
\begin{itemize}
\item Does the message directly address or reference children or use language aimed at a younger audience?
\item Does the message directly address or reference elderly community members, or discuss topics aimed at them?
\item Does the message introduce a degree of fear or feelings of uneasiness?
\item Is the message content complicated to understand?
\end{itemize}
\end{tcolorbox}
\end{quote}

\add {\subsection{Case Study}}
The \add{FABLE} framework aims to improve the ability for fact-checkers to evaluate the potential harm of an inaccurate or misleading claim and prioritize it as deserving of attention within their organizations. A FABLE analysis does not, however, result in a single, quantifiable ranking of urgency. Rather, the framework considers the many characteristics of misinformation that may lend themselves to harmful impact outlined under RQ1. It then clarifies the potential magnitude of harm based on \textit{variable} dimensions, while leaving open the possibility that, for strategic purposes, a fact-checker or organization may choose to focus on different categories of harm.

%For example, 
Harms that are more directed at individuals, such as doxxing, are different from those affecting broader society, such as health or election misinformation. How does one `calculate' that the first kind of categorical harm is less impactful than the second? 
% What happens, for example, when that first category becomes an issue of child sexual exploitation? 
When it comes to policies for handling certain categories of misinformation, excepting imminent physical danger, a fact-checker's or organizational approach is best handled through a mix of research, already defined policy, and community or expert consensus, rather than through any dynamic variables or individual judgment on the fly. Yet it may be possible to clarify which cases might be more urgent among similar cases, as well as to help distinguish where more strategic decision-making needs to take place. %This clarification can help both prioritization and impact assessment. 

%However, we note that categorical considerations still need to be accounted for on some occasions. For example, imminent physical harm has consistently been an important threshold regarding urgent response. 
Take, for example, two content scenarios:
\begin{itemize}
\item \textit{Scenario 1}: A post being shared on multiple platforms that claims that women who take a COVID-19 vaccine cannot get pregnant.
\item \textit{Scenario 2}: A post made by a political candidate accusing a rival candidate standing for political office of sexual assault.
\end{itemize}
Answering the questions connected to the framework, a fact-checker may determine that neither scenario seems likely to indicate high \textit{actionability} --- there are no calls for action or direct coordination efforts, for example --- but both may rank highly in the \textit{likelihood of spread}, whether because of its presence on multiple platforms or a connection to a pending election event. However, the former scenario may tally higher in the area of \textit{believability}, depending on the tone, audience and formatting of the message, while the latter results in some concerns about \textit{social fragmentation} since it touches upon public institutions and democratic processes. Depending on how the fact-checker or an organization decides to weigh these dimensions, as well as depending how they approach categorical considerations such as the potential for broad physical impact (e.g., on female bodies) versus individual reputational and social harms, it becomes apparent that prioritization decisions between these two claims may differ. %Working through examples like these can help fact-checkers think through their desired impacts, and a more strategic approach to prioritization.  

\section{Discussion}
%- About the framework (paste in the stuff from the white paper section?)
%- How can this framework be used/built on?
 %    - Lessons/ideas for computational tools
%- Considering beyond fact-checkers to other groups (content moderators, community moderators, public health communicators)

%\section{Implications}
\add{We discuss the implications of our results for practitioners and for researchers.} 
\add{From the interviews, we learned that fact-checkers are clearly concerned about harm, and that a focus on misinformation impact can be a productive avenue for investigation.}
\add{Whether for practice or for research, defining the difference between categorical and variable dimensions better clarifies key decisions related to policy and organizational strategy. In other words, the prioritization of categorical harms, such as physical versus psychological damage, is one that has to be made by humans, and cannot be determined by automated solutions. This recognition, as expressed in the framework itself, leads to a number of implications for both practice and research. } 

\add{\textit{Implications for practitioners.} For practitioners in fact-checking organizations, the framework is one solution to the expressed needs by our interviewees for more systematic approaches to prioritization.  Going through the questionnaire with several examples of claims can help organizations decide which kinds of categorical and variable dimensions they want to prioritize, or what kinds of impact they want to have, thereby allowing them to customize the framework to their own needs.}

Our expert reviewers saw the benefits of more nuanced and structured assessment while using our framework. In fact, R2 mentioned that different dimensions could be weighted differently, and cautioned us against adding the scores of each dimension together to form a conclusion or make comparisons. There may be value in using the framework in a periodic way for organizations seeking to structure their decision-making and increase their internal transparency, an expressed desire mentioned under RQ2. \add{Moreover, R1 emphasized that since the framework provides a comprehensive breakdown of the various aspects of harm, it could be an invaluable tool for training new fact-checkers and editors.}
%We acknowledge that different dimensions could have different interpretations and different levels of urgency under different context. Therefore, instead of providing a final "urgency" score, our framework only provides an individual score for each dimension. How to interpret or make use of of these scores is left for the organizations (or individuals) to decide on their own.
However, the initial application of the framework is not quick. Though answering these questions becomes faster with practice, as discussed under RQ2, fact-checking organizations are limited in their time, capacity, and resources. \add{Finding efficient ways to apply this framework to daily prioritization remains an area to be further studied.}

\add{\textit{Implications for researchers.} The need for efficiency creates opportunities for researchers.
As in past research and this paper, the complexities of harm and misinformation are high, which points to an area where HCI solutions can support structured, thoughtful, and faster approaches by fact checkers.}

\add{The different dimensions of the framework are conceptual tools for fact-checkers to negotiate their own processes. They also create design opportunities that give more decision-making control back to the users/organizations: rather than dictating that a piece of misinformation is more urgent the more viral it is, for example, applying the taxonomy lends itself towards interesting interactive interfaces, such as providing metrics in dashboards and visualizations that qualitatively and quantitatively represent the impact of their work. Additionally, sociotechnical solutions that seek to integrate the taxonomy within a single fact-checking organization might result in platforms where team members can negotiate their processes or discuss their biases and impact as an organization.}   

\add{A separate clear area of work is the potential automation of the taxonomy towards a prioritization queue or as a filtering and triage system---desires expressed by interviewees in section 3.3. Automated misinformation prioritization has been challenging due to the complexities and nuances of misinformation harm. With the recent advancements in Large Language Models (LLMs), our framework offers a pathway to guide and fine-tune these models, enabling them to generate structured and explainable analyses of harm.} Most of the reviewers (R1, R2, and R4) pointed out that the framework could be immensely beneficial if developed into a future automated tool for the preliminary screening, filtering, and prioritization of harmful content. For example, R4 responded ``\textit{This would absolutely be useful. A major issue (if not THE major issue) with existing misinformation-detection tools is that they perform extremely poorly at prioritizing false claims.}'' 
And R3 asserted that ``\textit{having a tool to detect harmful content and develop filters could be a way to prevent from post-traumatic stress disorder for example}.''  \add{In addition, a new research agenda for what indices might be automated, such as actionability or likelihood of spread, are clear opportunities for saving time and effort.}

\section{Limitations and Future Work}
While our effort aligns with the importance that professional fact-checkers place upon the potential harms of misinformation, we learned through our interviews that this was not their only concern. They also negotiate the needs of competing stakeholders, as well as their deliverables for technology platforms. Our effort however only focuses on harm.  How these other priorities might be taken into account is an area for future research. 
In addition, as pointed out by our fact-checkers, not all claims are fact-checkable. Many comments or conspiracy theories with potential for harm cannot be verified or debunked. Tools that could separate and identify fact-checkable claims, or that could isolate what elements of a claim are able to check, could also be incorporated into a more comprehensive tool for fact-check prioritization.

We chose to focus upon professionalized fact-checking processes. However, developments in collaborative, crowdsourced contexts such as Twitter/X Community Notes will offer additional ways to consider the problem of prioritization which may be beneficial for professional contexts. We also did not interview platform content moderators or platform operators who also do work to triage and prioritize content for review. We chose a narrower scope of inquiry due to our greater ability to recruit fact-checkers and report out about their internal organizational processes compared to platform operators. However, we believe that while platform operators have considerations that may be different from fact-checkers (e.g., platforms may still need a plan for claims that are difficult to verify, platforms have greater access to data, resources, and software tools than fact-checking organizations), many aspects of our prioritization dimensions are still relevant to content moderation and curation efforts on platforms. Future work can focus more specifically on empirical understanding of platform needs regarding misinformation prioritization.

While we interviewed fact-checkers working in multiple languages and countries, these questions do lean upon current understandings grounded in primarily English-language research. Questions may need to be adjusted over time or adapted to particular country and language contexts.
A Ukrainian fact-checker reminded us of the inequality in the available resources between countries and languages. As we consider automated processes for the future, it is notable that most NLP models are trained on popular languages such as English and Spanish, while few tools are available for people that speaks a minor language such as Ukrainian. In future development of AI or NLP, we should be reminded that these computational tools need to be language inclusive.

Finally, we note that the final validation by our reviewers is not robust. Rather, this evaluation served as an initial confirmation of our findings; more investigation of the framework in practice, and its possible transformation, is work that is yet needed.

\section{Conclusion}
In this work, we examined how fact-checkers prioritize which claims to inspect for further investigation and publishing, and what tools may assist them in their efforts. We explored this through interviews with 23 fact-checkers around the world, clarifying what aspects of misinformation they considered to create urgency or importance. These often revolved around the potential for the claim to harm others. We also learn more about the processes behind fact-checking decisions and suggestions for tools that could help fact-checkers with them. To address the needs articulated by these fact-checkers and others, \add{as well as a gap in the framework literature}, we present \add{the FABLE Framework of Misinformation Harms: a five-dimension questionnaire} to help fact-checkers negotiate the priority of claims. This effort was further validated by additional interviews with expert fact-checkers.

\begin{acks}
We appreciate the input of our fact-checkers and experts over the past two years, as well as others who have given us feedback. In particular, we wish to thank Franziska Roesner, Kate Starbird, Aimee Rinehart, as well as members of the Center for an Informed Public at the University of Washington for their feedback. We very much thank key contributions of Skyler Hallinan and Alexandra Bornhoft to this paper as well, and the styling/editorial work of Nevin Thompson on the questionnaire. Early stage work on this concept was supported with a small award from the Credibility Coalition and a gift from Meta. A larger project funded by the National Science Foundation’s Convergence Accelerator program under Award No. 49100421C0037 supported later work on the paper.
\end{acks}

% \begin{acks}
% \end{acks}
\bibliographystyle{ACM-Reference-Format}
\bibliography{REFERENCES}

\appendix
\newpage\section{Misinformation Harms Questionnaire}
The \add{FABLE Framework of Misinformation Harms} aims to support fact checkers in their efforts to discuss and prioritize in a more strategic way. 

Its five dimensions can help to clarify urgency within a category of harmful misinformation, thereby helping response teams. These multiple dimensions offer a holistic approach that encourages the evaluation of issues that might be missed if organizations are always only focused on responding to the most urgent content.  

In this framework, the degree of harm is positively correlated with degree of urgency: more "Yes" answers suggest more urgency or potential harm. However, this assumption may not always hold. We also recognize that organizations have considerations such as strategic areas of focus or internal resources. Therefore, organizations should use adjust the framework according to their needs. This is why there is no exact determination of how many ``Yes'' answers are required before a situation is considered urgent; it is up to each organization to determine. Or, while this framework aims to be holistic, an organization's use of it may also focus on a specific subset of dimensions or questions.

\begin{center}
  \makebox[\textwidth]{\includegraphics[width=\textwidth]{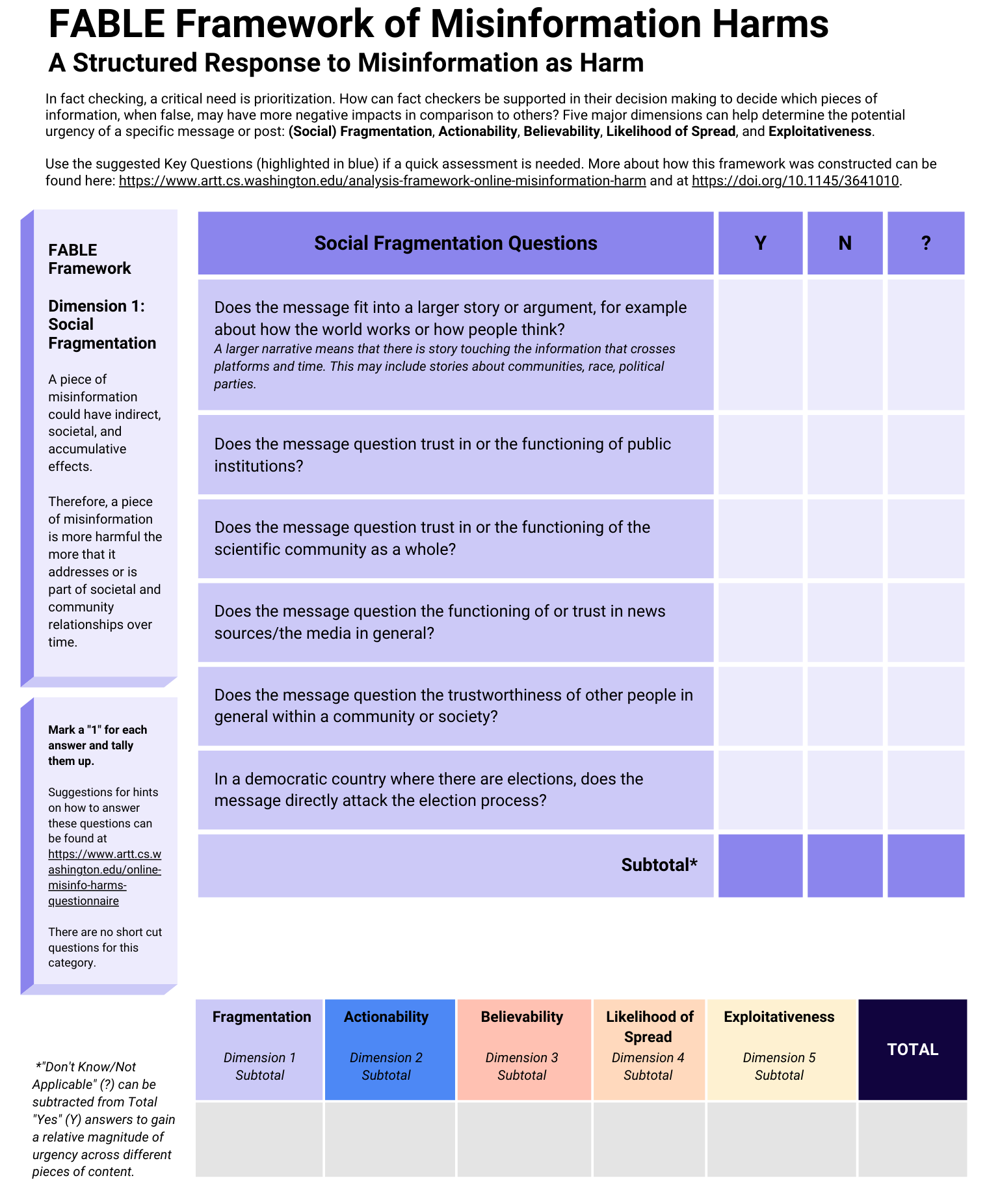}}
\end{center}
\newpage
\begin{center}
  \makebox[\textwidth]{\includegraphics[width=\textwidth]{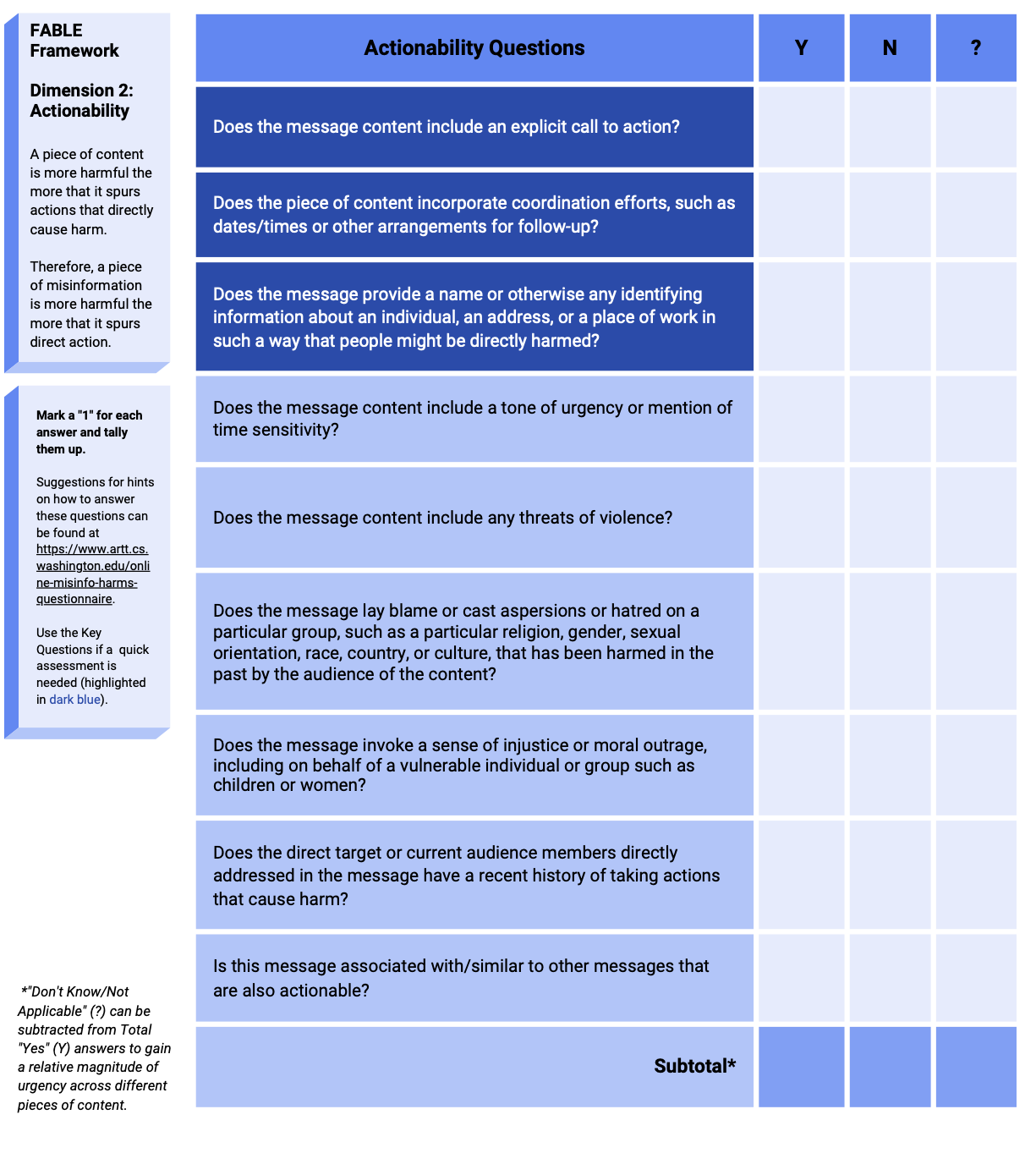}}
\end{center}
\newpage
\begin{center}
  \makebox[\textwidth]{\includegraphics[width=\textwidth]{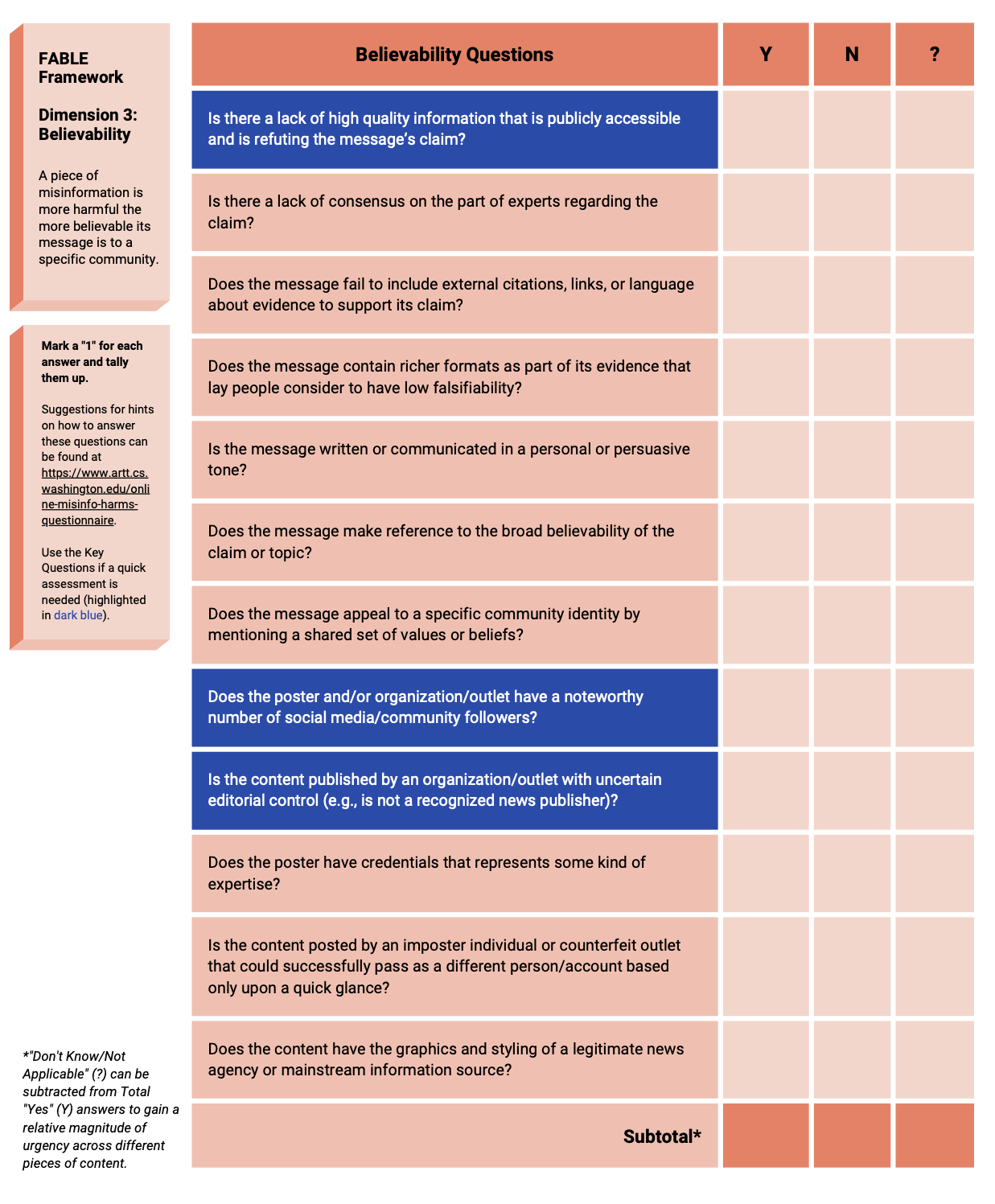}}
\end{center}
\newpage
\begin{center}
  \makebox[\textwidth]{\includegraphics[width=\textwidth]{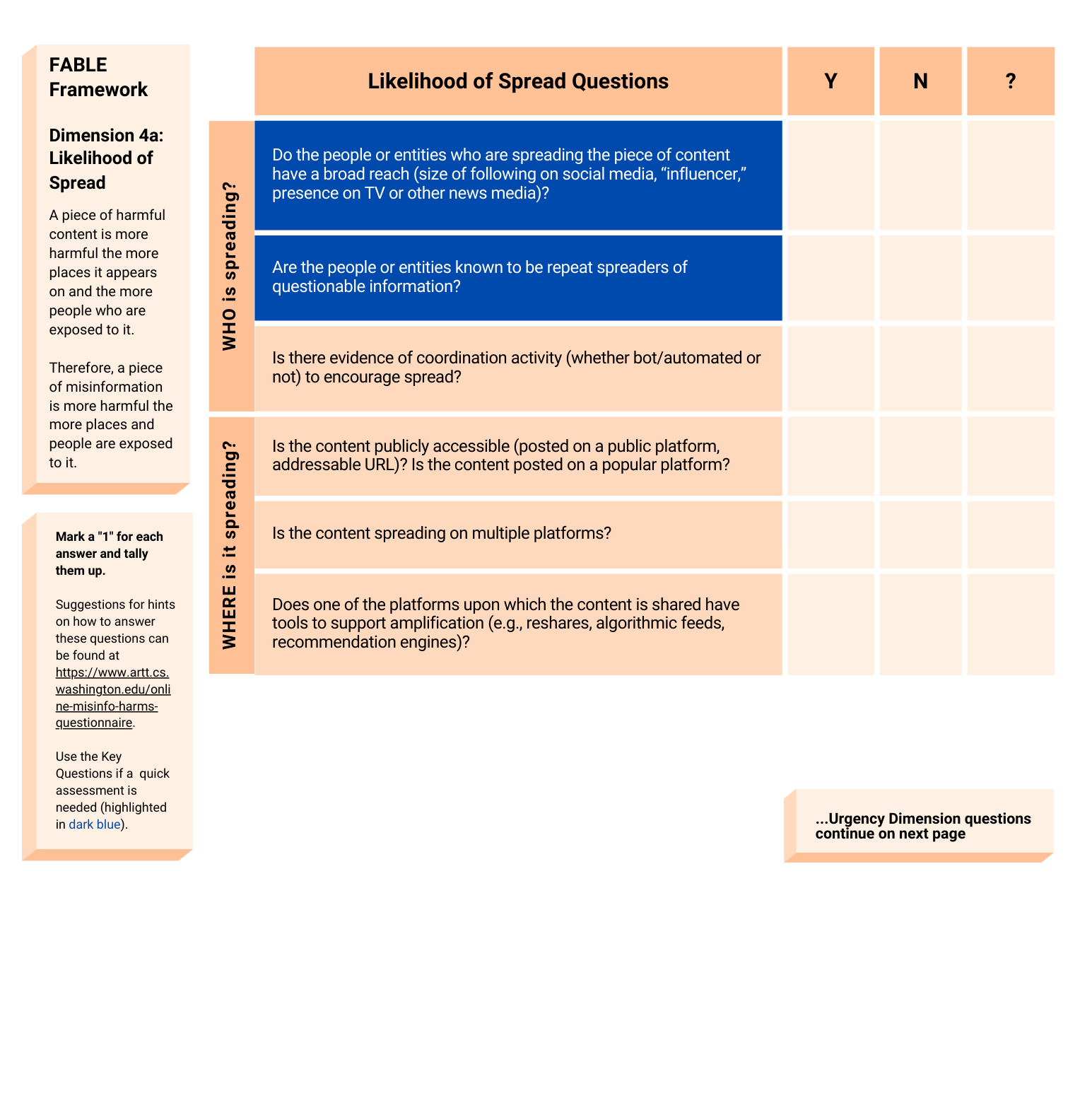}}
\end{center}
\newpage
\begin{center}
  \makebox[\textwidth]{\includegraphics[width=\textwidth]{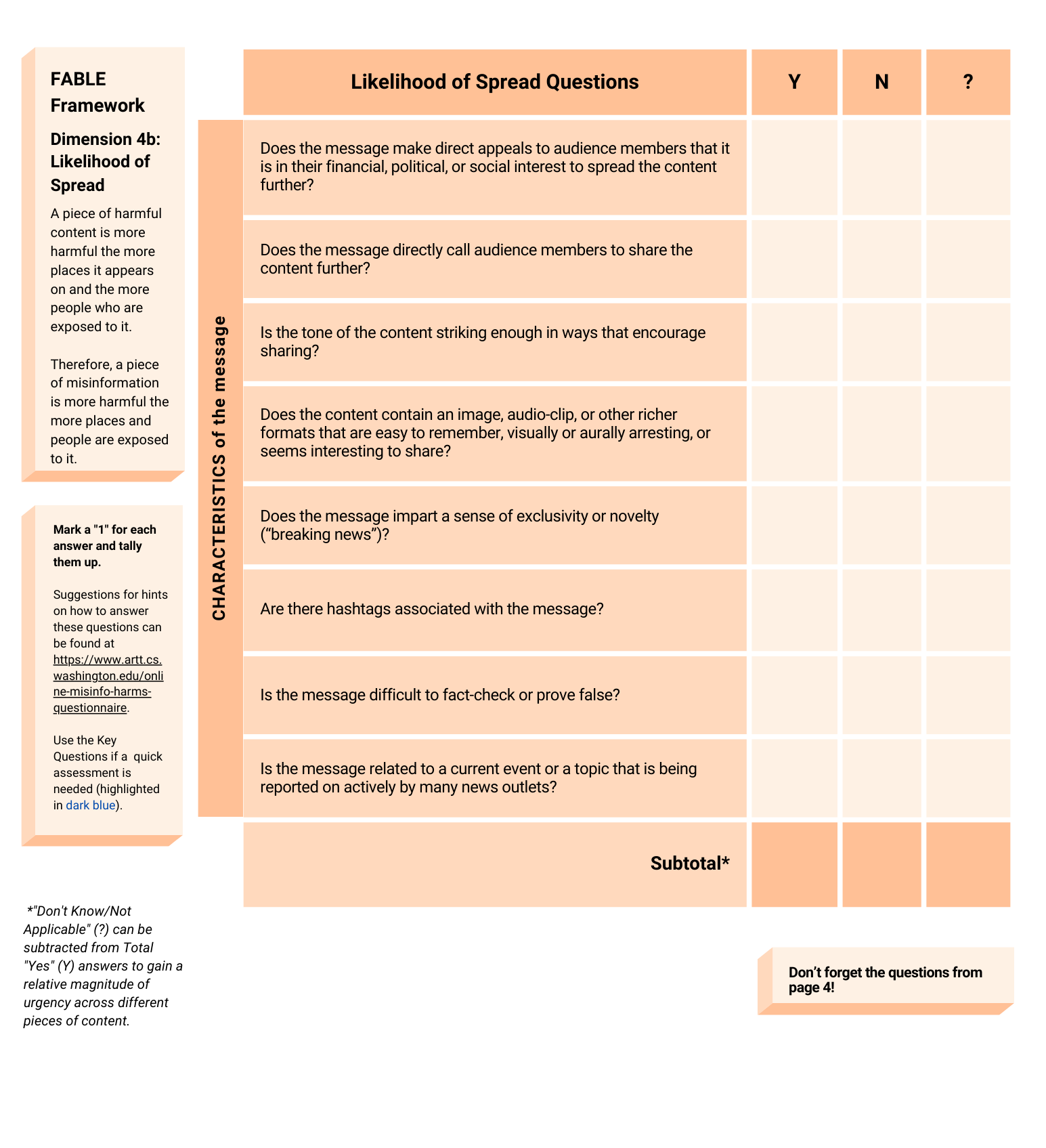}}
\end{center}
\newpage
\begin{center}
  \makebox[\textwidth]{\includegraphics[width=\textwidth]{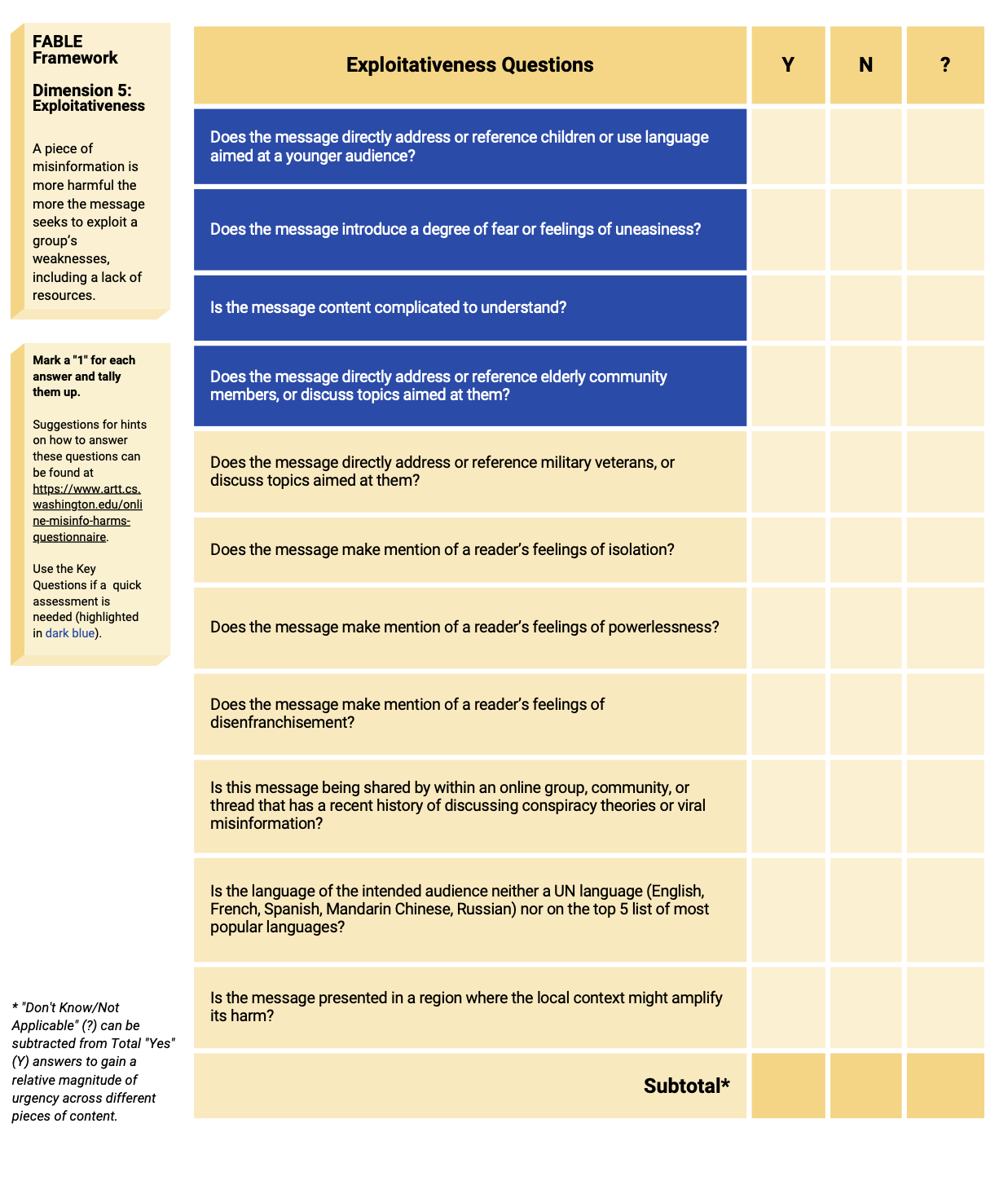}}
\end{center}

\end{document}